%% file: main.tex
\documentclass[conference]{IEEEtran}
\IEEEoverridecommandlockouts
\usepackage[utf8]{inputenc}
\usepackage[T1]{fontenc}
\usepackage[hyphens]{url}
\usepackage[breaklinks, colorlinks, linkcolor=black, anchorcolor=black, citecolor=black, urlcolor=black]{hyperref}
\usepackage{amsmath,graphicx,makecell}
\usepackage{enumitem} 
\graphicspath{{./fig/}}
\usepackage{threeparttable}
\usepackage{booktabs}

\usepackage{algorithm}
\usepackage{algorithmicx}
\usepackage{algpseudocode}

\usepackage{adjustbox}
\usepackage{booktabs}
\usepackage{amssymb}
\usepackage{wrapfig}
\usepackage{multicol}         
\usepackage{multirow}         
\usepackage{url}
\usepackage{tabularx}
\usepackage{xurl}
\usepackage{cite}
\usepackage{amsmath,amssymb,amsfonts,amsthm}
\usepackage{xcolor}
\usepackage{soul} 
\usepackage{graphicx}
\usepackage{textcomp}
\usepackage{url}
\usepackage{tabularx}
\usepackage{xurl}
\usepackage{threeparttable}
\usepackage{xcolor}
\usepackage{caption}
\usepackage{subcaption}
\usepackage{amsmath,amssymb}

\definecolor{midnightblue}{rgb}{0.1, 0.1, 0.44}

\newcommand{\+}[1]{\ensuremath{\boldsymbol{#1}}}

\usepackage{color}

\newcommand{\bluetext}[1]{{\leavevmode\color{blue}#1}}

\def\BibTeX{{\rm B\kern-.05em{\sc i\kern-.025em b}\kern-.08em
    T\kern-.1667em\lower.7ex\hbox{E}\kern-.125emX}}
\begin{document}

\title{An Active Learning-based Approach for Hosting Capacity Analysis in Distribution Systems}

\author{Kiyeob~Lee$^\ast$,~Peng~Zhao,~Anirban~Bhattacharya,~Bani~K.~Mallick~and~Le~Xie$^{\ddagger}$\\
        \{$^\ast$\IEEEmembership{Member},
        $^\ddagger$\IEEEmembership{Fellow}\},~IEEE
\thanks{Kiyeob Lee and Le Xie are with the Department of Electrical and Computer Engineering, Texas A\&M University, College Station, TX, USA.}
\thanks{Peng Zhao, Anirban Bhattacharya and Bani K. Mallick are with the Department of Statistics, Texas A\&M University, College Station, TX, USA.}
\thanks{$\ddagger$ Corresponding Author: Le Xie, email: le.xie@tamu.edu}
}

\maketitle

\begin{abstract}
With the increasing amount of distributed energy resources (DERs) integration, there is a significant need to model and analyze hosting capacity (HC) for future electric distribution grids. Hosting capacity analysis (HCA) examines the amount of DERs that can be safely integrated into the grid and is a challenging task in full generality because there are many possible integration of DERs in foresight. That is, there are numerous extreme points between feasible and infeasible sets. Moreover, HC depends on multiple factors such as (a) adoption patterns of DERs that depend on socio-economic behaviors and (b) how DERs are controlled and managed. These two factors are intrinsic to the problem space because not all integration of DERs may be centrally planned, and could largely change our understanding about HC. This paper addresses the research gap by capturing the two factors (a) and (b) in HCA and by identifying a few most insightful HC scenarios at the cost of domain knowledge. We propose a data-driven HCA framework and introduce active learning in HCA to effectively explore scenarios. Active learning in HCA and characteristics of HC with respect to the two factors (a) and (b) are illustrated in a 3-bus example. Next, detailed large-scale studies are proposed to understand the significance of (a) and (b). Our findings suggest that HC and its interpretations significantly change subject to the two factors (a) and (b).
\end{abstract}


\input{1-Introduction}
\input{2-Demand_Response}
\input{3-Interconnection}
\input{4-Formulation}

\input{9-Conclusion}
\Urlmuskip=0mu plus 1mu\relax  

\bibliographystyle{IEEEtran}
\bibliography{ref.bib}

\end{document}

%% file: 1-Introduction.tex
\section{Introduction}\label{sec:intro}

Power grids are an essential infrastructure of modern society and continue to expand as large-scale transitions towards sustainable and eco-friendly systems are expected. The trends are clear for the adoption of large-scale battery storage, plug-in electric vehicles (EVs), and renewable energy resources. Pathways towards these massive transitions definitely foster and accompany the creative destruction of technologies. Distributed generation (DG) is replacing the traditional greenhouse-gas emitting generation in an effort to move towards carbon-neutrality. Additionally, the transportation sector, where EVs are replacing the internal combustion engine cars, is getting electrified. The large-scale integration of distributed energy resources (DERs) such as DG, EVs, or battery energy storage systems will fundamentally reshape the future grid because DERs will gradually become integral parts of power systems. These movements bring up many challenges and opportunities that there are open questions about whether today's electric grid infrastructure is ready for large-scale transitions \cite{joskow2020transmission,xie2021toward}.

Along with large-scale transitions, today's electric grid is challenged to accommodate DERs as safely as possible towards a low-carbon economy. Notice that today's grid infrastructure was not built with DERs in mind and the electric grid will face unstable system states more frequently as the amount of DERs increases. It will eventually become unable to host additional DERs after a point without upgrading today's grid infrastructure. Hosting capacity analysis (HCA) addresses this line of questions by analyzing the amount of DERs that can be safely integrated into today's grid infrastructure. Understanding the hosting capacity (HC) is pivotal and particularly useful to distribution system operators because it allows them to anticipate and to prepare in advance for upcoming transitions.

There is a generic challenge in HCA that renders it a complex and interesting task. The integration of DERs involves a number of uncertain factors such as the location, capacity, and unpredictability. We underline that HCA investigates the system performance of today’s grid infrastructure with possible and upcoming integration of DERs without major infrastructure upgrades. Notice that today’s grid is a known realization, while the integration of DERs is an unknown forthcoming event. This leads to considering many scenarios (possibilities) for the integration of DERs. For instance, hosting a variable renewable energy source may be allowed in some locations without a major upgrade, but not in others. Clearly, the capacity, unpredictability, and intermittency of the generator also matter. Moreover, hosting DERs in a location may prevent other DERs from being integrated in other locations, which complicates the HCA in general.

The large-scale integration of DG may have adverse impacts on power system operations such as overvoltage, thermal overloading of the network, power loss, and system protection problems \cite{ebad2016approach,mohammadi2016challenges,gaunt2017voltage}. Studies have investigated the impacts of large-scale integration of EVs, and it has found that EVs can potentially cause thermal overloading and unexpected congestion in distribution systems \cite{leemput2014impact,linna2017congestion,johansson2019investigation}. HCA has been extensively studied for the integration of DG such as solar and wind generation \cite{baldenko2016determination,navarro2015probabilistic,abad2018probabilistic,fatima2020review} and for the integration of EVs \cite{bollen2017hosting,paudyal2021ev,rout2020hosting}. The integration of PV and wind power has been studied in \cite{liu2020probabilistic}, and there are recent papers that study the combined integration of PVs and EVs in \cite{fachrizal2021combined,e2022combined}. There are many optimization-based frameworks for HCA including \cite{madavan2022conditional,geng2021probabilistic}, and recently \cite{taheri2020fast} proposed an optimization-based HCA algorithm to accelerate computational time. Additionally, \cite{munikoti2022novel} removes the need to simulate a large number of scenarios using a spatial-temporal probabilistic framework.

\subsection{Potential Limitations of Conventional HCA}

Conventional HCA frameworks focus on maximizing the HC to identify the maximum amount of DERs that can be safely integrated into the grid. However, there are two other factors that are not typically accounted for in conventional HCA frameworks: (a) the adoption patterns of DERs, which depend on socio-economic and behavioral decisions, and (b) how DERs are controlled and managed. It is important to include these two factors in HCA frameworks, as they can significantly impact the performance of HCA.

By solving HCA as a maximization (optimization) problem, it identifies a global optimal solution and it could be a realistic global upper bound when a utility company can choose (and regulate) location and injection/consumption of all DERs. Notice standard centralized HCA frameworks are from a system operator or a central planner's perspective. However, actual adoption patterns of DERs may be far from the central planner's perspective when DERs are integrated and operate in a distributed manner rather than in a centralized manner. Then how much can we rely on conventional HCA frameworks when (a) and (b) play significant roles? To put the question differently, will adoption patterns of all DERs follow the solution of centralized HCA frameworks? Or will it start to deviate from the solution as non-centrally planned integration of DERs increases?

Moreover, standard centralized HCA frameworks including optimization-based HCA are designed to identify a global optimal solution (one extreme point between feasible and infeasible sets), i.e., one point with the largest sum of hosting capacities. Even if it can identify multiple extreme points, it is impossible to enumerate all extreme points in large-scale distribution grids because there are numerous extreme points. Then, how does the system operator deal with possibly numerous (hundreds, thousands or more) extreme points? It is unlikely that all extreme points are equally important. However, is it possible to scale down the number of extreme points and identify a few most insightful extreme points?



Due to these open questions, we propose an active learning-based HCA to include (a) and (b). Notice that the significance of the two factors will clearly vary in towns and cities. To the best of our knowledge, these questions have not been addressed.

\subsection{Contributions}

This paper begins with the following umbrella question: How can we model HCA more accurately and realistically? This question naturally leads us to investigate the way DERs are integrated and operate in distribution systems. DERs provide benefits to end-use customers who purchase them, and these customers drive the integration and operation of DERs. However, end-use customers are also influenced by socio-economic incentives and policies that impact their decisions. For example, an individual who is concerned about climate change may not be able to afford an EV due to its current cost, despite having a preference for consuming less fossil fuel. Similar stories can be found for other types of DERs. This work proposes an HCA framework to account for socio-economic incentives of end-use customers where customer preferences are abstracted into two factors (a) and (b). However, including (a) and (b) leads to a large number of scenarios. Thus, we propose a data-driven HCA framework and active learning algorithms. Here are the key contributions of our work:
\begin{itemize}
    \item We propose a data-driven HCA to include the two factors (a) and (b), active learning in HCA, and performance comparison of active learning in HCA.
    \item We present three case studies in HCA where the two factors (a) and (b) play significant roles: (1) adoption patterns of EVs concentrate and form clusters, (2) uncoordinated and coordinated EV charging, and 3) co-existence of PVs and EVs.
\end{itemize}

For factor (b), this work takes a data-driven approach to model how DER owners inject power into distribution systems using publicly available data sources for PVs and EVs. Although the dataset may not be universally applicable, the proposed methodology can be applied when high-fidelity data is collected by a utility company within their service territory.

For factor (a), we compared two possibilities: DERs located uniformly randomly over buses versus non-uniformly located over buses, where DERs are clustered in specific buses. We chose the latter option to capture clustering effects, which captures the intuitive idea of the rich gets richer phenomenon observed in social networks. In this context, it means that e.g., EV owners are more likely to have EV owner friends, or a wealthy local community is more likely to have more EV owners than a less wealthy local community. To understand the impact of clustering effects in HCA, we select clusters in the distribution system and exclusively located DERs in clusters. We compared two different clusters, one causing overvoltage problems and the other causing undervoltage problems, to highlight the importance of cluster location in factor (a) of HCA. This work demonstrates the importance of cluster location and factor (a) in HCA, despite the arbitrary selection of the EV cluster. Simulation results show that HC and its interpretations significantly change depending on factors (a) and (b).

The rest of this paper is organized as follows. Section \ref{sec:HCA} introduces a data-driven HCA, characteristics of HC and discussions about the importance to include (a) and (b) in HCA. Section \ref{sec:AL} introduces active learning and apply active learning in HCA. Section \ref{sec:case} starts with a 3-bus network example to illustrate active learning and characteristics of HC. Next, it presents, on a larger network, performance comparisons of active learning, and case studies that include two factors (a) and (b) in HCA. We conclude the article in Section \ref{sec:conc}.

%% file: 2-Demand_Response.tex
\section{Problem Formulation}\label{sec:HCA}

First, we present a DistFlow model to represent distribution grids which holds true if the distribution network is a tree \cite{low2014convex}. Then, we formulate HCA as a data-driven method using DistFlow model, and provide detailed discussions on hosting capacity with respect to the two factors (a) and (b).

\subsection{DistFlow Model}
Consider a distribution grid of radial networks $N = (V, E)$ with buses (vertices) $V$ and lines (edges) $E \subset V \times V$. Denote the cardinality of a set $X$ as $|X|$. Let nodal real and reactive injections be $\+p \in \mathbb{R}^{|V|}$ and $\+q \in \mathbb{R}^{|V|}$ respectively, nodal squared voltage magnitudes be $\+v \in \mathbb{R}^{|V|}$, real and reactive flows on lines be $\+P \in \mathbb{R}^{|E|}$ and $\+Q \in \mathbb{R}^{|E|}$ respectively. Then, DistFlow model \cite{baran1989network} of a distribution grid can be written as follows:
\begin{subequations}\label{eq:DistFlow}
    \begin{align}
        & p_{j}+P_{i j}=r_{i j} l_{i j}+\sum_{k:(j, k) \in E} P_{j k} \label{eq:DistFlow-a}\\
        & q_{j}+Q_{i j}=x_{i j} l_{i j}+\sum_{k:(j, k) \in E} Q_{j k} \label{eq:DistFlow-b}\\
        & v_{i}-v_{j}=2\left(r_{i j} P_{i j}+x_{i j} Q_{i j}\right)+\left(r_{i j}^{2}+x_{i j}^{2}\right) l_{i j} \label{eq:DistFlow-c}\\
        & l_{i j}=\frac{P_{i j}^{2}+Q_{i j}^{2}}{v_{i}} \label{eq:DistFlow-d}\\
        & i \in V \cup\{0\}, ~ j \in V, ~~ \forall ~ (i, j) \in E. \nonumber
    \end{align}
\end{subequations}
where $r_{ij}$ and $x_{ij}$ are resistance and reactance of line $(i,j) \in E$ respectively. Constraints represent real/reactive power flow \eqref{eq:DistFlow-a}-\eqref{eq:DistFlow-b} and voltage limit along the line \eqref{eq:DistFlow-c}-\eqref{eq:DistFlow-d}. DistFlow model characterizes the relationship among the above variables associated with the network \cite{low2014convex}. We denote the model \eqref{eq:DistFlow} as
\begin{align*}
[\+P,\+Q,\+v] = \operatorname{DistFlow}(\+p,\+q).
\end{align*}

\subsection{Hosting Capacity Analysis}

Given a DistFlow model of radial networks $N = (V, E)$, we are interested in hosting capacity (HC), i.e., the amount of DERs that can be safely integrated into today's grid without violating or upgrading the infrastructure. In the existing grid infrastructure, assume that there are daily baseline real/reactivate load profiles $\{\+d[t],\+e[t]\}_{t \in T}$ over time $T$ before any integration of DERs . Suppose a candidate location of DERs is denoted as $L \subset V$. A scenario denoted as $\+\psi \in \mathbb{R}^{|L|}$ represents a set of DERs installed and a net profiles $\{ \+\alpha[t] \}_{t \in T}$ (net injection or consumption) over locations $L$ for time $T$. Now, consider the following feasibility problem:

\begin{subequations}\label{eq:feasibility}
\begin{align}
& {[\+{P}[t], \+{Q}[t], \+{v}[t]]=\operatorname{DistFlow}(\+{p}[t], \+{q}[t])} && t \in [T], \label{eq:feasibility-df} \\
& \+{p}[t]=\+{A}^{L} \operatorname{diag}(\+{\alpha}[t]) \+\psi-\+{d}[t] && t \in [T], \label{eq:feasibility-pf1} \\
& \+{q}[t]=\+{A}^{L} \operatorname{diag}(\+\eta)\operatorname{diag}(\+{\alpha}[t]) \+\psi-\+{e}[t] && t \in [T], \label{eq:feasibility-pf2} \\
& \underline{\+{v}} \leq \+{v}[t] \leq \overline{\+{v}} && t \in [T], \label{eq:feasibility-volt} \\
& \left(P_{i j}[t]\right)^{2}+\left(Q_{i j}[t]\right)^{2} \leq\left(\bar{S}_{i j}\right)^{2} && \forall ~ (i, j) \in E, t \in [T], \label{eq:feasibility-line}
\end{align}
\end{subequations}
where $[T] := \{1, 2, \ldots, T \}$. Constraints represent nodal real/reactive power balance \eqref{eq:feasibility-pf1}-\eqref{eq:feasibility-pf2}, voltage magnitude \eqref{eq:feasibility-volt} and line flow limits \eqref{eq:feasibility-line}. Here, matrix $\+A^{L} \in \{0,1\}^{|V|\times |L|}$ is the DER location-to-bus adjacency matrix, i.e., $A^{L}_{ij} = 1$ if the $j^{th}$ DER is at bus $i$ and $A^{L}_{ij} = 0$ otherwise. We assume that DERs operate in the maximum power point tracking (MPPT) mode and maintain fixed power factor $\eta \in \mathbb{R}^{|L|}$ by simple reactive power control.

Given a feasibility problem \eqref{eq:feasibility} and a set of scenarios $\{ \+\psi \}$, we propose HCA as a feasibility problem as follows. In \eqref{eq:feasibility}, a scenario $\+\psi$ is either feasible or infeasible for each $t \in T$ and denote it as $\mathcal{O}_{t}(\psi) = 1$ if feasible and $\mathcal{O}_{t}(\psi) = 0$ otherwise. Finally, define the evaluation of a scenario $\+\psi$:
\begin{align}\label{eq:HC_SA}
    \mathcal{O}(\+\psi) = \begin{cases}
    1 ~~ \text{if} ~~ \frac{1}{T} \sum_{t=1}^{T} \mathcal{O}_{t}(\+\psi) \ge \bar{\epsilon} \\
    0 ~~ \text{otherwise}
    \end{cases}
\end{align}
where $\bar{\epsilon} \in [0, 1]$ determines how many time steps it is allowed to be infeasible out of $T$. It is a computationally tractable approximation (sample average approximation) of more general chance-constrained problems \cite{geng2019data}.

Observe that the feasibility problem \eqref{eq:feasibility} includes 1) baseline real/reactive load profiles $\{ \+{d}[t], \+{e}[t] \}_{t \in T}$, 2) a scenario $\+\psi \in \mathbb{R}^{|L|}$ and 3) generation/fload profiles $\{ \+{\alpha}[t] \}_{t \in T}$ of a scenario $\+\psi$. In other words, the feasibility problem \eqref{eq:feasibility} determines whether the scenario $\+\psi$ (and its corresponding profiles $\{ \+{\alpha}[t] \}_{t \in T}$) is feasible or infeasible, and a set of scenarios $\{\+\psi\}$ is evaluated on the same grid. Note that the system performance should comply with acceptable standards such as all nodal voltages within a range from $0.95$ p.u. to $1.05$ p.u. \eqref{eq:feasibility-volt} and line loading less than $100\%$ of normal ampere rating \eqref{eq:feasibility-line}.

Clearly, it is expected that HC will be smaller if $\bar{\epsilon} = 1$ and that HC will be larger if $\bar{\epsilon} < 1$. For example, suppose a scenario $\+\psi$ has to satisfy $\frac{1}{T} \sum_{t=1}^{T} \mathcal{O}_{t}(\+\psi) = \Bar{\epsilon} = 1$, i.e., $\mathcal{O}_{t}(\+\psi) = 1$ for all $t$. Then, the scenario $\+\psi$ needs to be conservative to be feasible for all $t$. On the other hand, if $\bar{\epsilon}$ is not tight, for example, $\bar{\epsilon} = 0.9$, then HC is larger yet it raises concerns about system security. We investigate the empirical relationship between HC and $\bar{\epsilon}$ in case studies (Section \ref{sec:case}).


\subsection{Characteristics of Hosting Capacity}

\begin{figure}[t]
    \centering
    \includegraphics[width=0.75\linewidth]{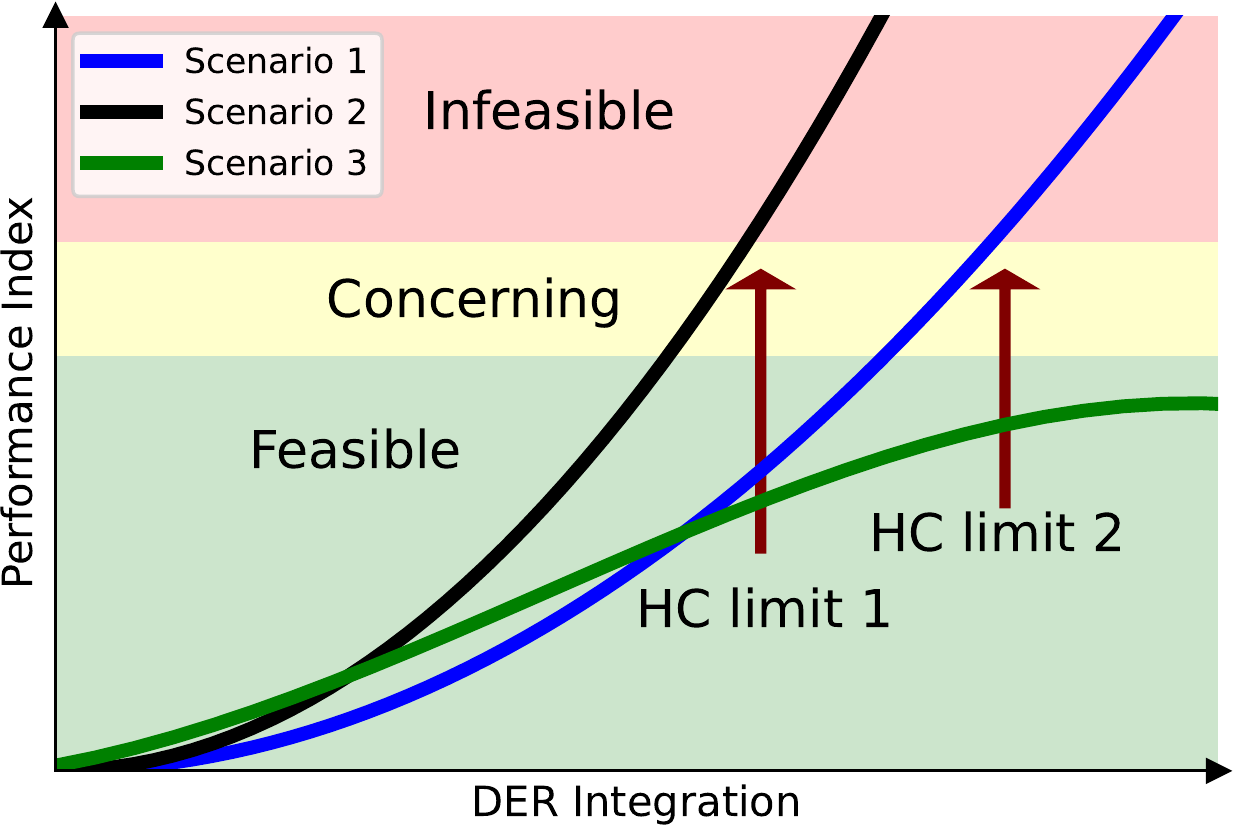}
    \caption{Characteristics of Hosting Capacity}
    \label{fig:HC_concept}
\end{figure}

The major purpose of HCA is to understand how feasible and infeasible sets are separated and extreme points (HC limits) between the two sets. Accurate formulation of HCA needs to achieve the two goals. Fig. \ref{fig:HC_concept} presents a schematic illustration that there could be numerous HC limits between feasible and infeasible sets. To support this illustration, we provide a 3-bus network example in simulation results.


Note that a standard optimization-based HCA is designed to identify a global optimal solution (i.e., one point with the largest sum of hosting capacities), which is one extreme point between feasible and infeasible sets. In HCA, we argue below that identifying multiple extreme points is more worthwhile and illustrative than identifying one global optimal point. Recently, \cite{madavan2022conditional} proposed a conditional value at risk (CVaR) based approach that identifies multiple feasible scenarios.

Notice that even if a method can identify multiple extreme points (or scenarios), there are numerous extreme points and it may be impossible to enumerate all the points in large-scale distribution grids. Recall that extreme points (and scenarios between the two sets) correspond to a possible integration of DERs. This presents a challenge because it is unclear how to deal with the numerous extreme points. There could be hundreds, thousands or more extreme points, but there are remaining open questions: are all extreme points equally important to consider? Or can we scale down and identify a few most insightful extreme points? We hold these questions and raise another pivotal aspect.


On the other hand, we claim that HC depends on two factors and they are crucial in HCA frameworks: (a) adoption patterns of DERs driven by socio-economic behavior and (b) how DERs are controlled and managed. We provide detailed discussions and examples why it is important to include (a) and (b) below.

Regarding (a), EVs can exist at any location after installing a charging station in principle, however, EVs are more quickly adopted in wealthy towns than in others. Similarly, small-scale PVs can exist at any location, but adoption patterns are driven by socio-economic and behavioral decisions. Thus, it is clearly expected that adoption patterns of DERs are not going to be uniformly random in terms of location, but are driven by socio-economic and behavioral decisions correlated with household location \cite{muratori2018impact}. Notice that adoption patterns of DERs may significantly vary in towns and cities. Regarding (b), even if we know where DERs exist and concentrate, it is equally important to know how DERs are controlled and managed. For example, it is easily predicted that the grid can host more EVs when they are coordinated whereas the grid can host less EVs when they are uncoordinated. Still, there is an open question about the degree of coordination of EVs, which may greatly vary in towns and cities.

Although it may be impossible to precisely know about (a) and (b) without domain knowledge such as transportation and demographic information, observe that the two factors are intrinsic to DERs because DERs are distributed by definition. However, when additional domain knowledge is available, observe that including (a) and (b) in HCA scales down extreme points to consider. Thus, we argue that it may be necessary to include (a) and (b) in HCA in great detail. Yet, there are a large number of scenarios and, hence, active learning is proposed.

%% file: 3-Interconnection.tex
\section{Active Learning in Hosting Capacity Analysis}\label{sec:AL}
In this section, we introduce active learning frameworks, active learning algorithms, and apply to HCA.

\subsection{Introduction to Active Learning}

Many machine learning applications include a case that there is a set of unlabeled data instances from a domain $\mathcal{X}$. Each instance $x \in \mathcal{X}$ has an unknown label from a label set $\mathcal{Y}$ and $x$ can be queried by an oracle. The objective of active learning is maximize performance while minimizing expensive data labeling. More specifically, the goal is to train a classifier, a function $h: \mathcal{X} \rightarrow \mathcal{Y}$ without making too many queries to an oracle. An oracle denoted as $\mathcal{O}$, for example, can be a human annotator or a computational resource which maps an input to a label, i.e., $\mathcal{O} : \mathcal{X} \rightarrow \mathcal{Y}$ where $\mathcal{X}$ is an input space and $\mathcal{Y}$ is a label space.

\begin{figure}[ht]
  \centering
  \includegraphics[width=0.8\linewidth]{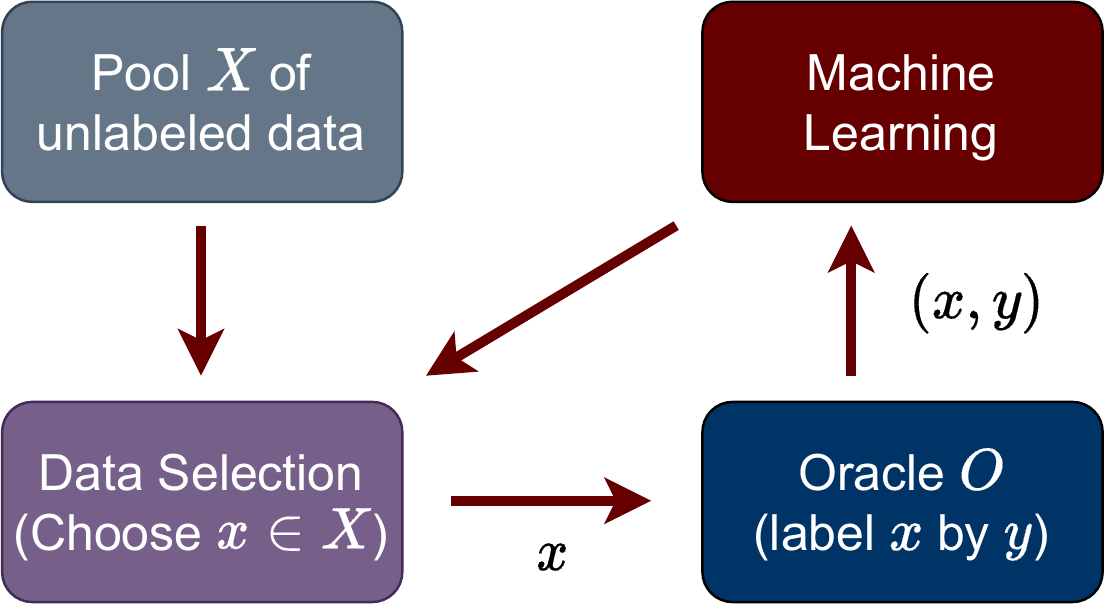}
  \caption{Pool-based Active Learning Framework}
  \label{fig:AL_framework}
\end{figure}

There are three standard frameworks in active learning literature \cite{settles2009active} and we first introduce a framework called pool-based active learning illustrated in Fig. \ref{fig:AL_framework}. To begin, there is a pool (set) of unlabeled data instances denoted as $\mathcal{X}$. A unique property of active learning is to query unlabeled data instances to be labeled by an oracle $\mathcal{O}$, and there is an associated cost to access the oracle $\mathcal{O}$ in terms of monetary value or computational resources. Therefore, it is essential to effectively query which unlabeled data instances in $\mathcal{X}$ to be labeled by an oracle $\mathcal{O}$ to improve performance and does not exceed a cost budget (i.e., do not make too many queries). For this purpose, define a function called query strategy in which its goal is to effectively select unlabeled data instances to be labeled and we discuss about query strategy in detail below. When a query strategy selects $x \in \mathcal{X}$ to be labeled, the oracle $\mathcal{O}$ annotates a label $y$. Then a pair $(x, y)$ is added to the dataset for training a machine learning model, and this cycle is repeated until a budget is exhausted. This process is essentially outlined in Fig. \ref{fig:AL_framework}.

We quickly highlight distinctions between three frameworks of active learning \cite{settles2009active}. They are different based on how unlabeled pool $\mathcal{X}$ is accessed, while the other pieces (query strategy, oracle, machine learning) remain mostly the same: 1) Pool-based has access to a pool of unlabeled data to begin; 2) Stream-based has access to an unlabeled data instance at a time rather than a pool $\mathcal{X}$ and, after labeling/rejecting it, another instance arrives; and 3) Synthesis-based has no access to unlabeled data at all and it is allowed to generate unlabeled data instances. While three active learning frameworks exist, we mark that the distinction between the frameworks is not very significant to HCA because they only differ on how an unlabeled pool $\mathcal{X}$ is accessed. Due to this, only the pool-based active learning framework is considered in this work.

We now introduce three query strategies, commonly used in active learning. First, define $\mathbb{P}(y|x;\theta)$ the probability that model predicts a label $y$ provided that input is $x$ under model parameter $\theta$. Entropy strategy is defined as $\phi^{Ent}(x) := - \sum_{y \in \mathcal{Y}} \mathbb{P}(y|x;\theta) \log \mathbb{P}(y|x;\theta)$. Information density strategy is defined as $\phi^{ID}(x) = \phi^{Ent}(x)(\frac{1}{|\mathcal{X}|}\sum_{x' \in \mathcal{X}} \frac{x^{T}x'}{\|x\| \| x'\|})^{\beta}$ where $\beta$ is a parameter that controls the relative importance of the density term. Lastly, define a uniformly random strategy $\phi^{U}(x) = \frac{1}{|\mathcal{X}|}$, i.e., all elements $x \in \mathcal{X}$ are equally likely to be selected. Observe that $x^{*} \in \arg\max_{x \in \mathcal{X}} \phi^{Ent}(x)$ is a maximizer of the entropy (information); $x^{*} \in \arg\max_{x \in \mathcal{X}} \phi^{ID}(x)$ is a maximizer of the information density; Uniformly random strategy $\phi^{U}(\cdot)$ is a Monte Carlo method, also called passive learning in active learning literature.

Entropy strategy $\phi^{Ent}$ and information density strategy $\phi^{ID}$ generally outperform uniformly random sampling (Monte Carlo methods), although the best strategy is application-dependent \cite{settles2009active}. On the other hand, performance of active learning depends on the pool $\mathcal{X}$ that $\phi^{Ent}$ and $\phi^{ID}$ suffer when there are too many outliers in $\mathcal{X}$ \cite{karamcheti2021mind}. This may not be surprising because entropy and information density strategies are noise-seeking in a sense to maximize the information gain in $\mathcal{X}$. It is also clear that $\phi^{Ent}$ and $\phi^{ID}$ suffer when scenarios in $\mathcal{X}$ are combinatorial. Similarly, performance of active learning increases when the number of outliers decreases in the pool. Thus, it is advised that the choice of query strategy $\phi$, $\mathcal{X}$ and machine learning model will all affect the performance of active learning in practice.

We are ready to present a pool-based active learning Algorithm \ref{algo:AL}. It starts with the following inputs: 1) labeled set $\mathcal{L}$, 2) unlabeled set $\mathcal{X}$, 3) query strategy $\phi$, 4) query size $B \in \mathbb{N}$ and 5) computational budget $K \in \mathbb{N}$. The labeled set $\mathcal{L}$ can be an empty set without loss of generality, a pool (set) of unlabeled data instances $\mathcal{X}$ is labeled by oracle $\mathcal{O}$, and both query size and budget are positive integers. In line 5, it takes a maximizer $x_{b}$ of query strategy $\phi(\cdot)$ (e.g., that maximizes information) and access to the oracle $\mathcal{O}$ to get a label $y_{b}$ in line 6. After labelling, append the pair $(x_{b},y_{b})$ to the labeled set $\mathcal{L}$ and remove $x_{b}$ from the unlabeled pool $\mathcal{X}$ (line 7). Algorithm \ref{algo:AL} repeats lines 5-7 in a for loop (line 4) and use the labeled set $\mathcal{L}$ for training (line 3). Finally, Algorithm \ref{algo:AL} repeats lines 3-7 until the maximum budget $K$ is reached, and returns $\theta$ and $\mathcal{L}$.

\begin{algorithm}
\caption{Pool-based Active Learning}\label{algo:AL}
\begin{algorithmic}[1]
\State Input : Labeled set $\mathcal{L}$, unlabeled pool $\mathcal{X}$, query strategy $\phi$, query size $B$, budget $K$
\For {$k=0,1,\ldots, K$}
  \State $\theta = \text{train}(\mathcal{L})$ 
  \For {$b=0,1,\ldots, B$}
    \State $x_{b} = \arg \max_{x \in {\mathcal{X}}} \phi(x)$
    \State $y_{b} = \mathcal{O}(x_{b})$ 
    \State $\mathcal{L} := \mathcal{L} \cup (x_{b},y_{b}) ~ \text{and} ~ \mathcal{X} := \mathcal{X} \setminus \{ x_{b} \}$
\State \textbf{Return} $\theta, \mathcal{L}$
    \EndFor
\EndFor
\end{algorithmic}
\end{algorithm}


\subsection{Active Learning in Hosting Capacity Analysis}

We now apply active learning to HCA. In Fig. \ref{fig:AL_framework}, recall that there are four major components and we explain all components in detail in terms of HCA.

First, it is necessary to have a set (pool) $\mathcal{X}$ of unlabeled scenarios $\mathcal{X} = \{ \+\psi \}$ in the active learning. In practice, $\mathcal{X}$ may not be available, and it is important to generate it. In the next section, we provide explanations on how to generate a set of scenarios in detail. It is worth highlighting that evaluation of one scenario $\+\psi$ in power system simulation software may be computationally inexpensive (e.g., in the order of seconds). However, the total number of scenarios could be immense and, thus, the total computational time may be unacceptable. For example, suppose we are interested in installing 5 solar PVs and there are 50 candidate locations (buses). This already tells that there are more than $10^6$ ( $>>$ 50 Choose 5) PV scenarios. Similarly, the number of scenarios for possible EV adoptions could be too gigantic. Thus, it is necessary to compose a set of scenarios $\mathcal{X}$ with domain experts' knowledge and evaluate a subset of scenarios in $\mathcal{X}$ selectively, not exhaustively, to determine the HC within a computational budget; otherwise, the total number of scenarios may be beyond a reasonable computation budget. We provide our suggestions in detail in case studies where we propose simple ways to design $\mathcal{X}$.

Next, there is a data selection component where its goal is to select a scenario $\+\psi \in \mathcal{X}$ such that $\arg\max_{\+\psi \in \mathcal{X}} \phi$ for a given query strategy. After choosing a scenario to evaluate, the oracle evaluates the feasibility of the selected scenario $\+\psi$, which is taken to be OpenDSS \cite{OpenDSS} in this work, and give a label (feasible or infeasible). In general, the oracle can be any software that evaluates the feasibility/infeasibility of a scenario by solving power flow equations \eqref{eq:HC_SA}. Lastly, there is a machine learning component and it updates the query strategy $\phi$ by using machine learning algorithms so that data selection component effectively selects an unlabeled scenarios to evaluate.

In this work, active learning Algorithm \ref{algo:AL} and machine learning tasks are implemented in Python programming language, and interface with OpenDSS through a Python package called OpenDSSDirect \cite{OpenDSSDirect}.

%% file: 4-Formulation.tex
\section{Simulation Results}\label{sec:case}

The proposed active learning-based HCA is evaluated. First, a simple example of a 3-bus network is provided. Next, on a 123-bus network, we provide performance comparison of active learning algorithms and case studies to illustrate the importance of (a) and (b) in HCA.

\subsection{3-bus Network}

Matpower \cite{zimmerman2010matpower} provides a number of distribution system test cases, and we modify the 4-bus network by removing one bus and use it in this work. Scenarios are uniformly randomly generated, i.e., $(\+\psi_{1}, \+\psi_{2}) \in [0,4]^{2}$ p.u., to illustrate the active learning and characteristics of HC. Three algorithms are proposed and modified from Algorithm \ref{algo:AL} as follows: 1) a Monte Carlo strategy $\phi^{U}$ in line 5 and there is no need for training in line 3, 2) an entropy strategy $\phi^{Ent}$ in line 5 and two layer fully-connected neural networks in line 3 and 3) an information density strategy $\phi^{ID}$ in line 5 and two layer fully-connected neural networks in line 3. Details about fully-connected neural networks are referred to \cite{shalev2014understanding}.

\begin{figure}[hbt!]
\centering
\begin{minipage}{.05\linewidth}
\centering
\rotatebox{90}{Monte Carlo}
\end{minipage}\hfill
\begin{minipage}{.03\linewidth}
\centering
\rotatebox{90}{$\+\psi_{1}$}
\end{minipage}\hfill
\begin{minipage}{.285\linewidth}
\centering
\textbf{50 scenarios}
\includegraphics[width=1\linewidth]{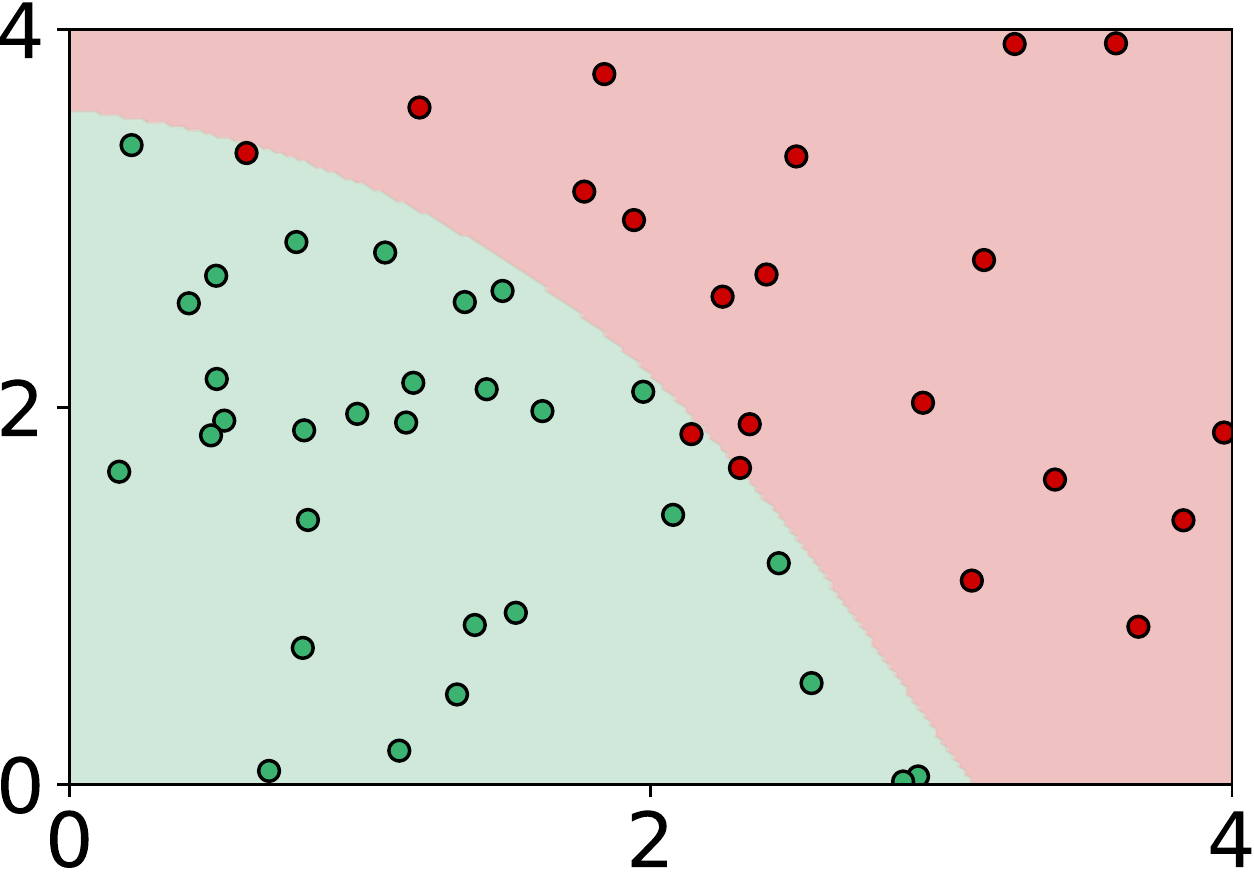}
$\+\psi_{2}$
\end{minipage}\hfill
\begin{minipage}{.03\linewidth}
\centering
\rotatebox{90}{$\+\psi_{1}$}
\end{minipage}\hfill
\begin{minipage}{.285\linewidth}
\centering
\textbf{100 scenarios}
\includegraphics[width=1\linewidth]{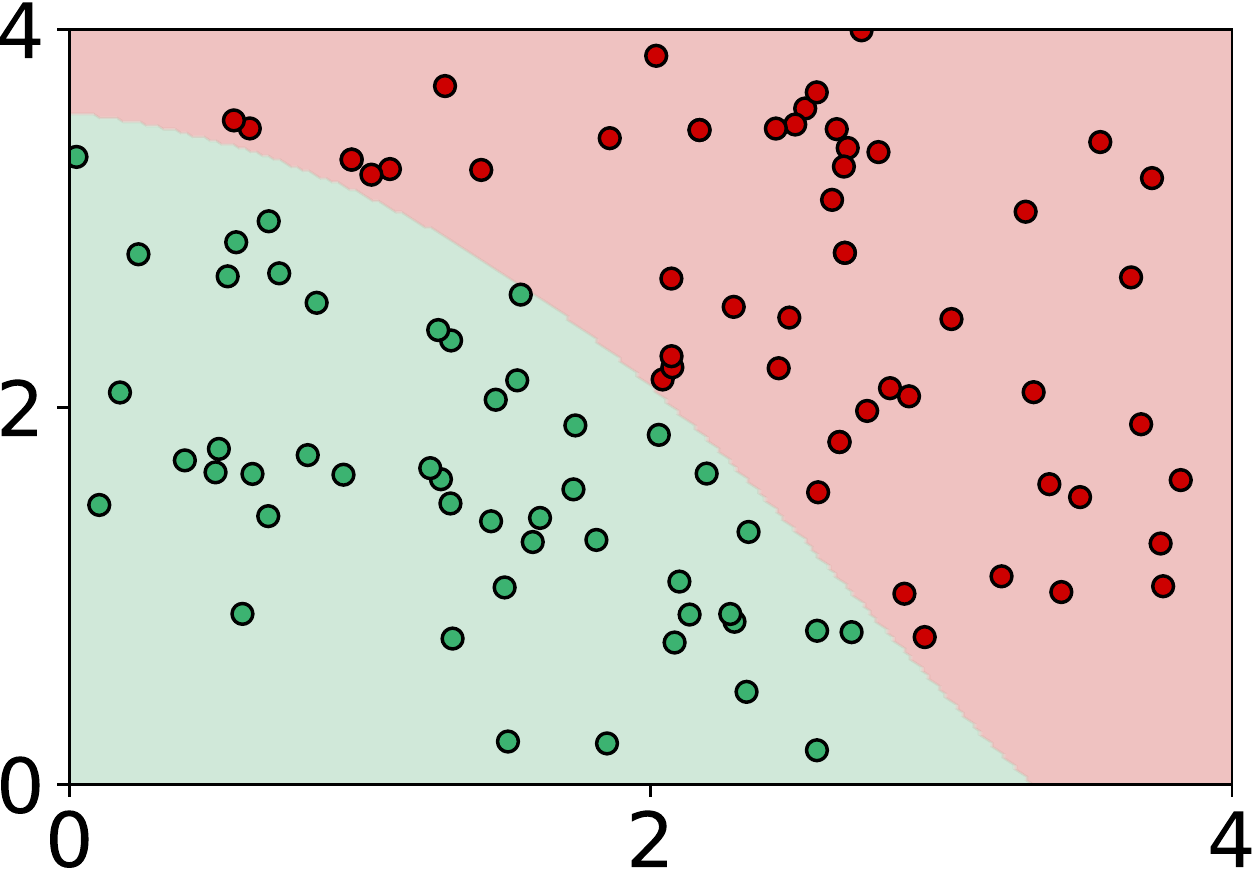}
$\+\psi_{2}$
\end{minipage}\hfill
\begin{minipage}{.03\linewidth}
\centering
\rotatebox{90}{$\+\psi_{1}$}
\end{minipage}\hfill
\begin{minipage}{.285\linewidth}
\centering
\textbf{500 scenarios}
\includegraphics[width=1\linewidth]{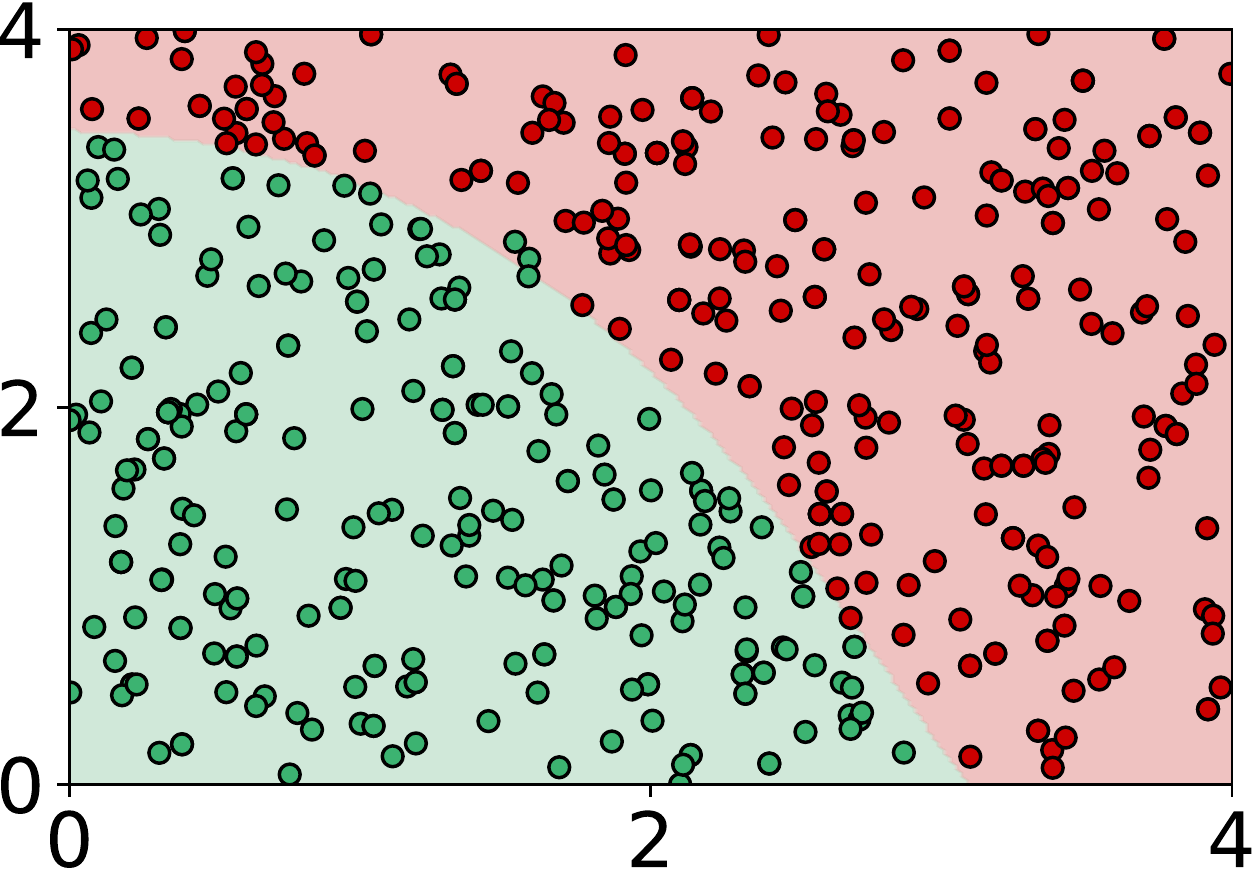}
$\+\psi_{2}$
\end{minipage}\hfill
\centering
\begin{minipage}{.05\linewidth}
\rotatebox{90}{Entropy}
\end{minipage}\hfill
\begin{minipage}{.03\linewidth}
\centering
\rotatebox{90}{$\+\psi_{1}$}
\end{minipage}\hfill
\begin{minipage}{.285\linewidth}
\centering
\includegraphics[width=1\linewidth]{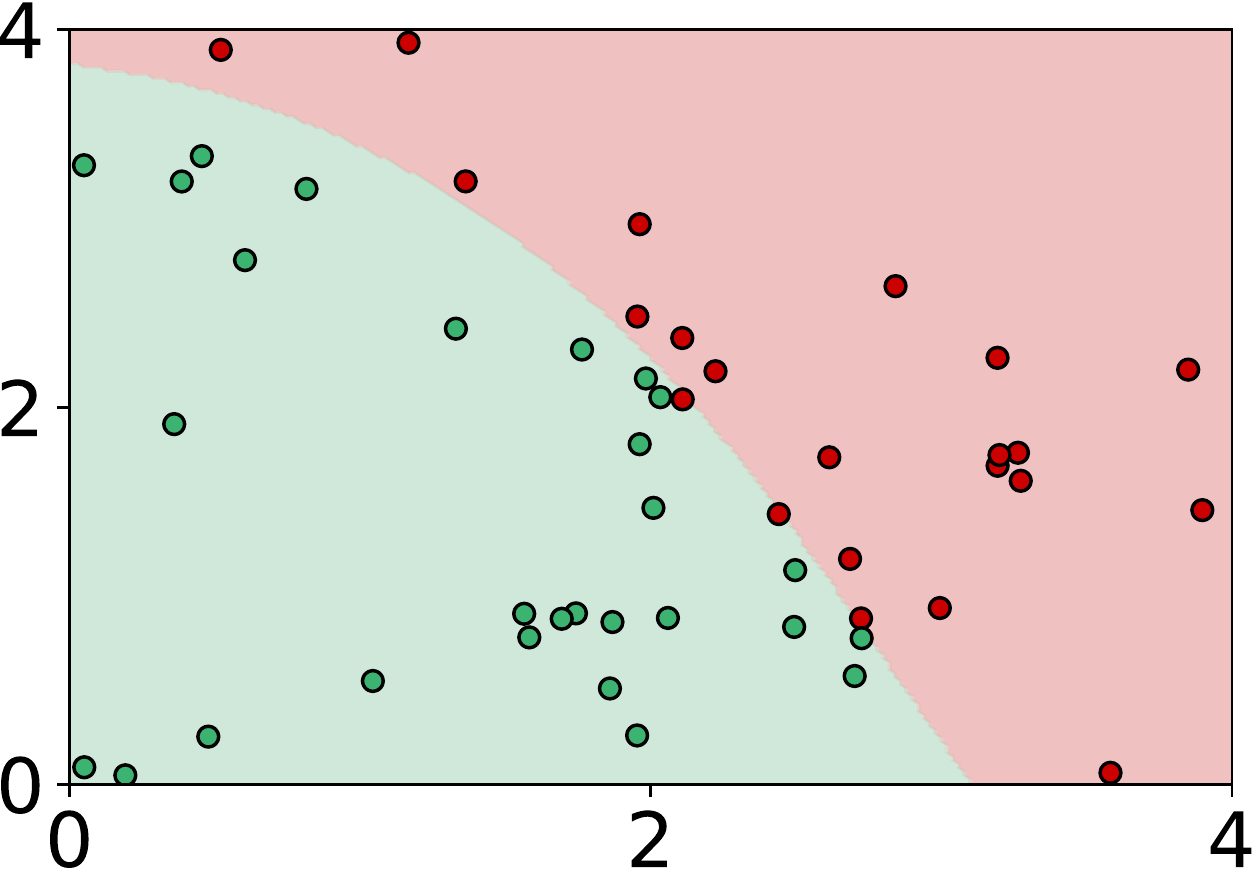}
$\+\psi_{2}$
\end{minipage}\hfill
\begin{minipage}{.03\linewidth}
\centering
\rotatebox{90}{$\+\psi_{1}$}
\end{minipage}\hfill
\begin{minipage}{.285\linewidth}
\centering
\includegraphics[width=1\linewidth]{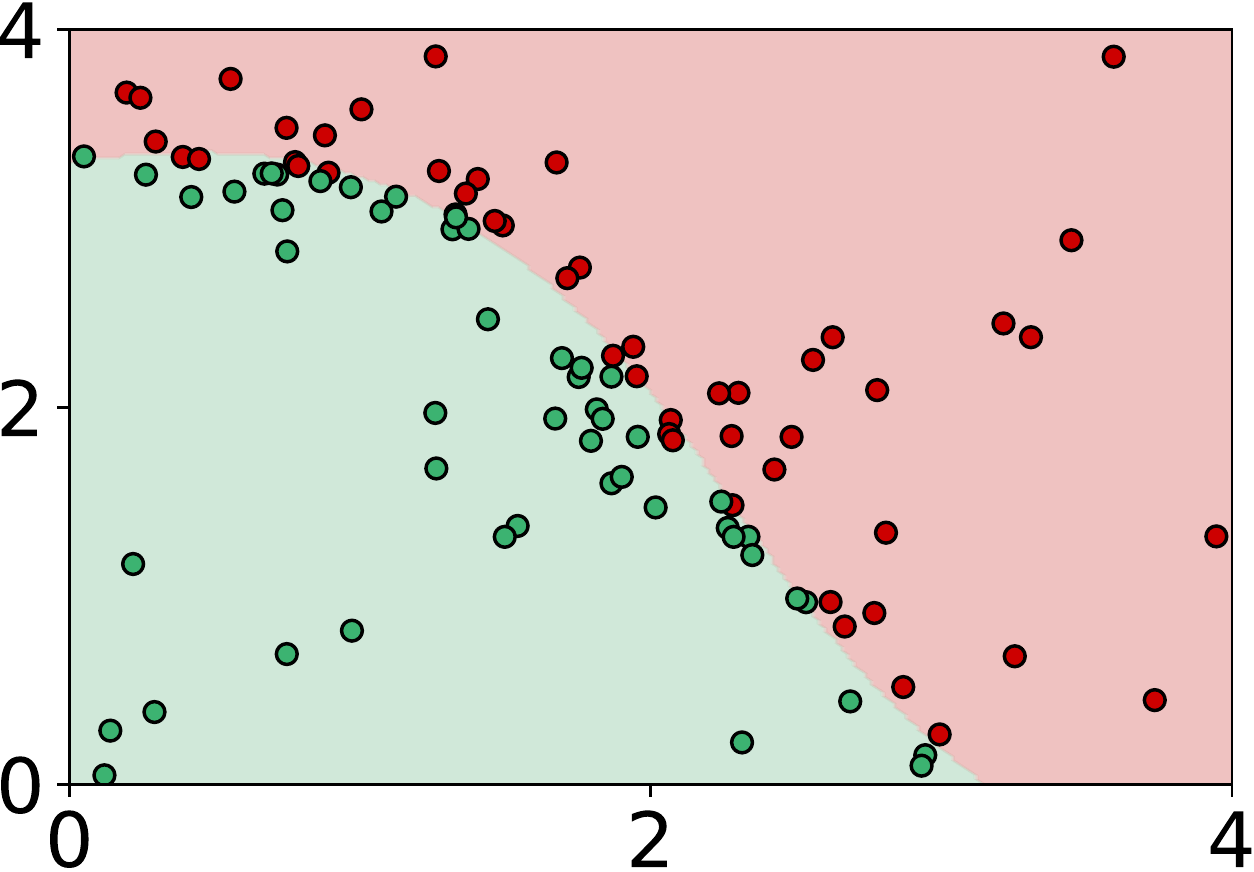}
$\+\psi_{2}$
\end{minipage}\hfill
\begin{minipage}{.03\linewidth}
\centering
\rotatebox{90}{$\+\psi_{1}$}
\end{minipage}\hfill
\begin{minipage}{.285\linewidth}
\centering
\includegraphics[width=1\linewidth]{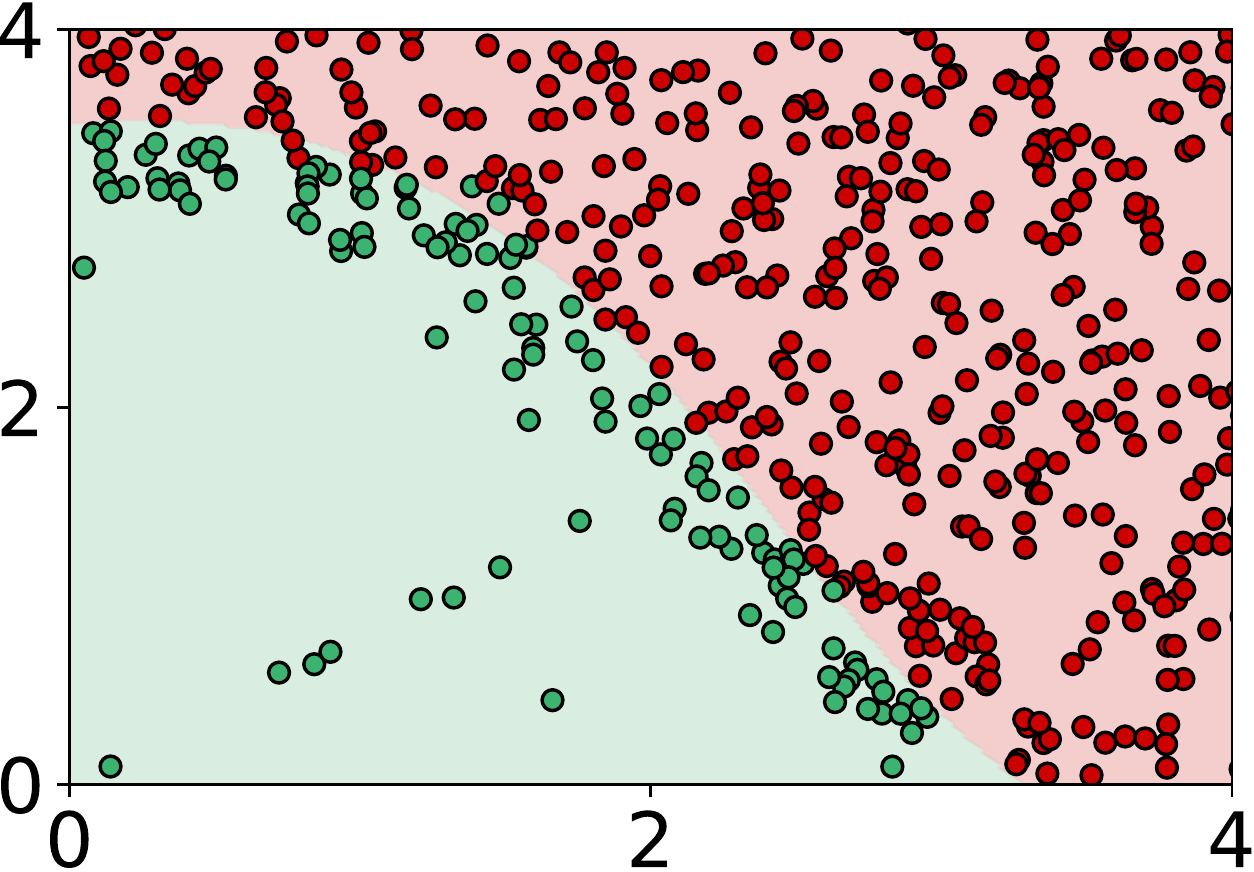}
$\+\psi_{2}$
\end{minipage}\hfill
\centering
\begin{minipage}{.05\linewidth}
\rotatebox{90}{Information Density}
\end{minipage}\hfill
\begin{minipage}{.03\linewidth}
\centering
\rotatebox{90}{$\+\psi_{1}$}
\end{minipage}\hfill
\begin{minipage}{.285\linewidth}
\centering
\includegraphics[width=1\linewidth]{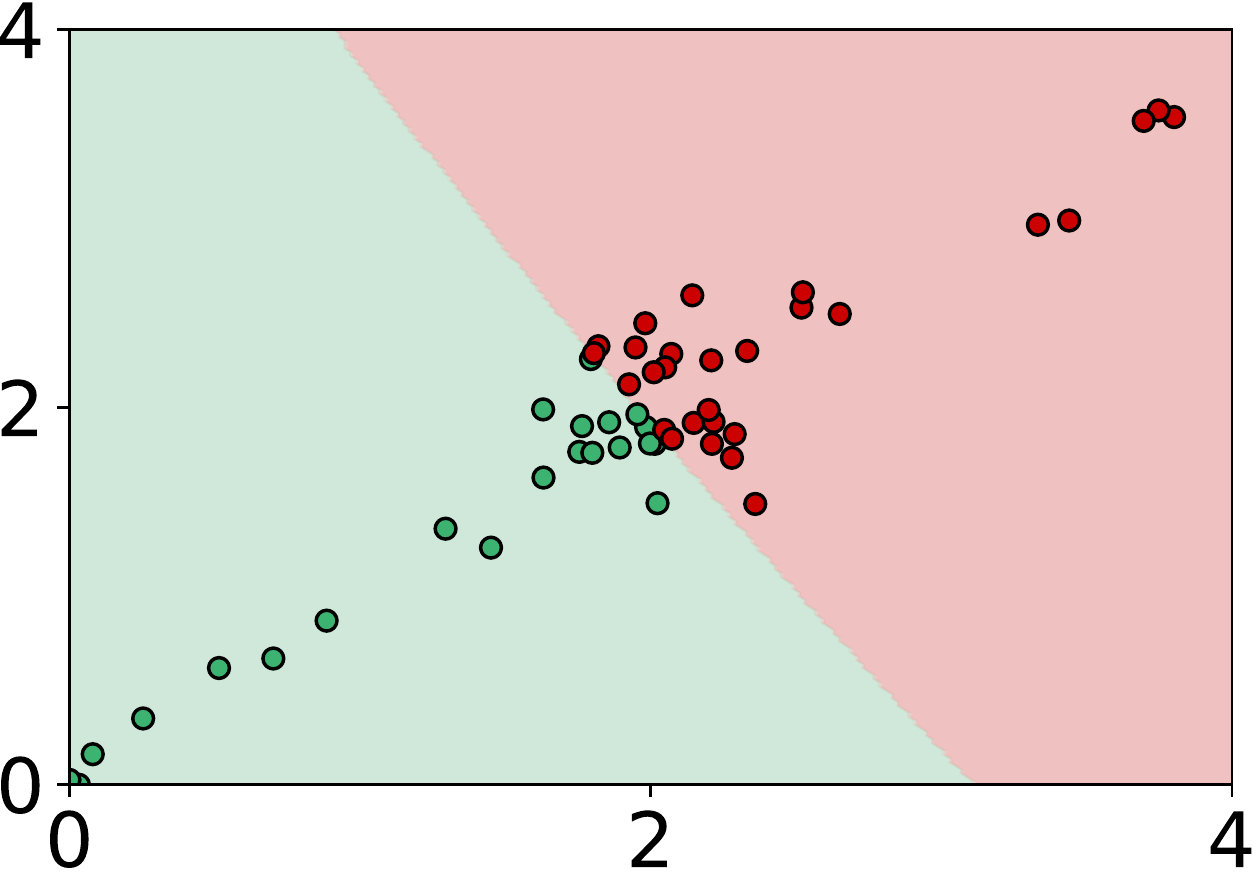}
$\+\psi_{2}$
\end{minipage}\hfill
\begin{minipage}{.03\linewidth}
\centering
\rotatebox{90}{$\+\psi_{1}$}
\end{minipage}\hfill
\begin{minipage}{.285\linewidth}
\centering
\includegraphics[width=1\linewidth]{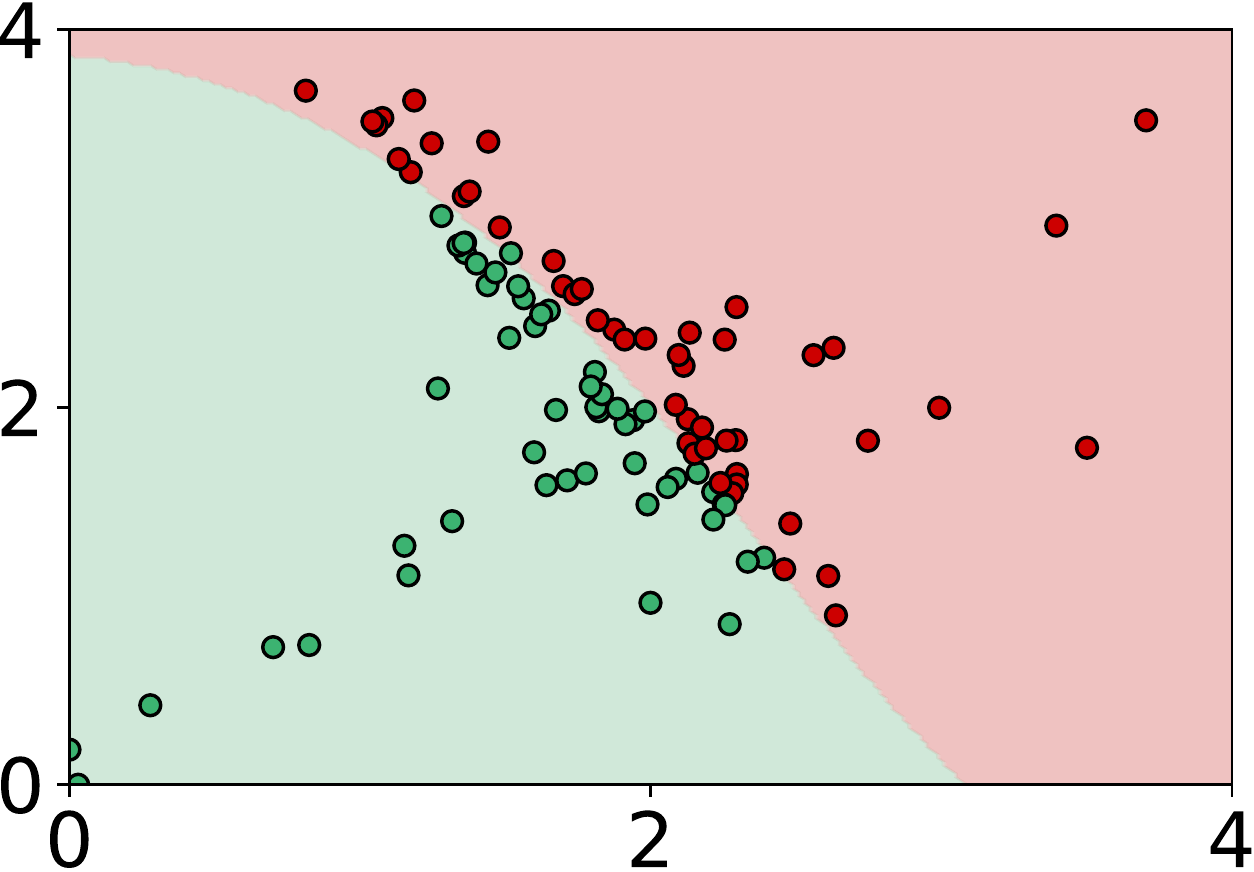}
$\+\psi_{2}$
\end{minipage}\hfill
\begin{minipage}{.03\linewidth}
\centering
\rotatebox{90}{$\+\psi_{1}$}
\end{minipage}\hfill
\begin{minipage}{.285\linewidth}
\centering
\includegraphics[width=1\linewidth]{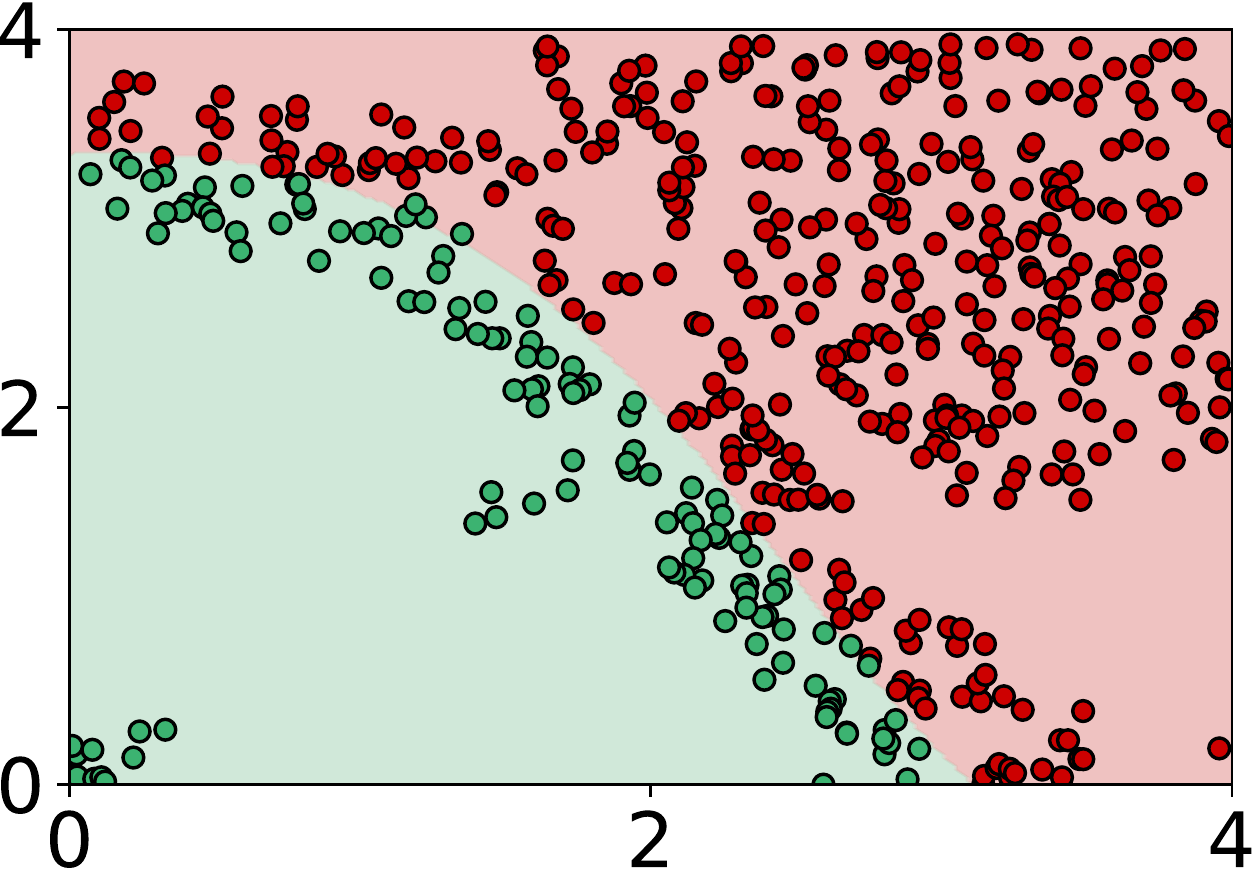}
$\+\psi_{2}$
\end{minipage}\hfill
\caption{Visualization of scenarios $\{\+\psi\}$ after evaluating 50, 100 and 500 scenarios on the 3-bus network example with Monte Carlo Strategy in the top, Entropy Query Strategy in the middle, and Information Density Query Strategy in the bottom. Green and red points represent feasible and infeasible scenarios respectively. Shaded regions represent a separating hyperplane between two sets of data points.}
\label{fig:3-bus}
\end{figure}

Active learning is illustrated in Fig. \ref{fig:3-bus} where it shows how the sampling process evolves in three active learning algorithms. 50, 100, 500 scenarios are evaluated and placed, where green and red points represent feasible and infeasible scenarios respectively. Also, two sets of feasible and infeasible scenarios are separated by support vector machines (SVMs) \cite{shalev2014understanding}. Notice that 50 and 100 scenarios are especially contrasted where entropy and information density strategies exhibit noticeable concentration around the boundary and where Monte Carlo strategy does not exhibit significant concentration. This is predicted because of the noise-seeking property of $\phi^{Ent}$ and $\phi^{ID}$, and note that it could be an advantage or disadvantage depending on applications. Also, notice that information density strategy exhibits more narrow concentration compared to entropy strategy.

In this 3-bus example, observe that there are numerous scenarios close to the boundary between feasible and infeasible scenarios where they range from the top left corner to the bottom right corner in Fig. \ref{fig:3-bus}. Moreover, the maximum hosting capacity (the largest sum of hosting capacities $\psi_1$ and $\psi_2$) is simply one of them and is far from characterizing the boundary. The existence of numerous extreme points complicates our understanding about the HC especially for more practical large-scale distribution grids because of two reasons: 1) it is unclear which extreme points are useful from a system operator's perspective in foresight; 2) it is not straightforward to visualize the boundary. On the other hand, when additional domain knowledge about (a) and (b) is available, it is possible to identify a few most insightful extreme points at the cost of (a) and (b) and not to worry about all others. For large-scale distribution grids, this is particularly advantageous because it is impossible to enumerate all extreme points in general.


Lastly, notice that there is an implicit assumption in this example that scenarios are uniformly generated in $[0,4]^2$. In Euclidean space, it is straightforward to choose a proper domain ($[0,4]^2$) by visualization. However, in higher dimensions, it is not clear how to choose a proper domain (e.g., $[0,4]^n$) because it is not easy to visualize. Therefore, choosing a proper domain involves a design problem irrespective of scenarios as well as (a) and (b).

\subsection{123-bus Network and Scenario Setup}

We take the IEEE 123 Node Test Feeder as our choice of a distribution grid and take a default baseline nodal demand profile, i.e., we are given a default real/reactive load profiles $\{\+d[t],\+e[t]\}_{t \in T}$ over $T = 144 ( = 24 * 6)$ time steps for a day. In the 123 bus Feeder, there are 91 load buses.

A set of scenarios $\mathcal{X} = \{ \+\psi \}$ is based on the two data sources \cite{muratori2018impact,lew2013western} to capture two factors (a) adoption patterns of DERs that depend on socio-economic behaviors and (b) how DERs are controlled and managed from.

We take EV consumption profiles proposed in \cite{muratori2018impact} where it models residential power demand including EV charging of a group of households. Residential power demand profiles are 10-min resolution for 365 days with two different light-duty EV charging power levels typically used in the United States: Level 1 and Level 2. According to SAE J1772 standard, Level 1 charger operates at 1.92kW and Level 2 charger can operate at up to 19.2kW depending on the charging station. In this work, we assume that all EV customers are equipped with Level 1 chargers and take one day out of 365 days to match the baseline nodal demand profiles. Similarly, we take solar PV generation profiles proposed in \cite{lew2013western} that solar data consists of 1 year of 5-minute resolution with 60 PV plants. We modified the time-scale of solar data accordingly to match the time-scale of other data. While dataset \cite{muratori2018impact, lew2013western} provides generation and consumption profiles of DERs, they are decoupled to distribution grids and there is no information about locations of DERs. Thus, they are meaningful only if we can properly choose locations of DERs in distribution grids and generate scenarios accordingly.

\subsection{Performance of Active Learning Algorithms}

\textbf{Design of Scenarios}: A set of scenarios $\mathcal{X}_{1}$ is designed to serve as a baseline for performance comparison as follows. Assume that EVs are located in all load buses and that the number of EVs (positive integer) in each bus is drawn according to the Poisson distribution with a parameter $\lambda$ and that there are $N_{\text{EV\_TYPE}} = 20$ residential EV charging (consumption) behaviors \cite{muratori2018impact}. Similarly, assume that PV plants are allowed to be located in a pre-selected 30 buses and that the number of PV plants $N_{\text{PV}}$ is drawn uniformly randomly in each bus and that there are $20$ PV generation (injection) behaviors \cite{lew2013western}. When the location and net injection of DERs are chosen over all buses, it defines a scenario $\+\psi$. Notice that a scenario characterizes the two factors (a) and (b). That is, (a) in terms of choosing a number of DERs in locations, and (b) in terms of net injection, by appropriately changing the parameters in a scenario.

A set of scenarios $\mathcal{X}_{1} = \{ \+\psi \}$ is generated by incrementally increasing $\lambda$ from 1.0 to 1.5 and $N_{\text{PV}}$ from 10 to 30. We change the two parameters $\lambda$ and $N_{\text{PV}}$ to have scenarios lie in Feasible, Concerning, and Infeasible sets in Fig. \ref{fig:HC_concept}. The purpose is to have scenarios (extreme points) between feasible and infeasible sets. We generate $\mathcal{X}_{1}$ has a cardinality $|\mathcal{X}_{1}| \approx 16,000$ as describe above. All scenarios in $\mathcal{X}_{1}$ are evaluated once to obtain the maximum possible hosting capacity within $\mathcal{X}_{1}$. Lastly, fix the probabilistic lower bounds $\bar{\epsilon} = 0.98$ in Eq. \ref{eq:HC_SA} and report that about $53\%$ of the scenarios in $\mathcal{X}_{1}$ is feasible while $47\%$ of the remaining scenarios is infeasible.

\begin{figure}[hbt!]
\begin{minipage}[t]{.49\linewidth}
\centering
\includegraphics[width=1\linewidth]{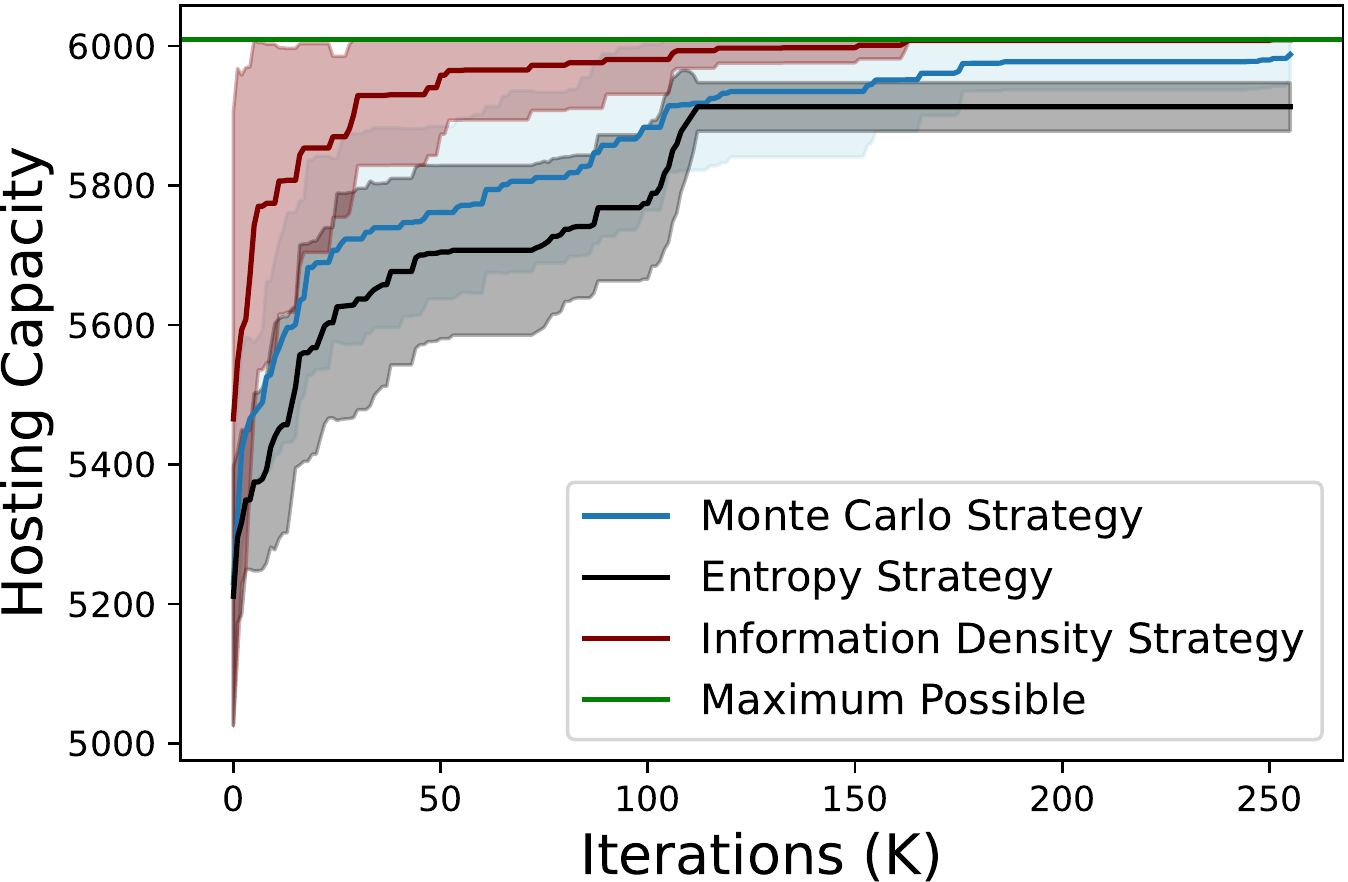}
(a)
\end{minipage}\hfill
\begin{minipage}[t]{.49\linewidth}
\centering
\includegraphics[width=1\linewidth]{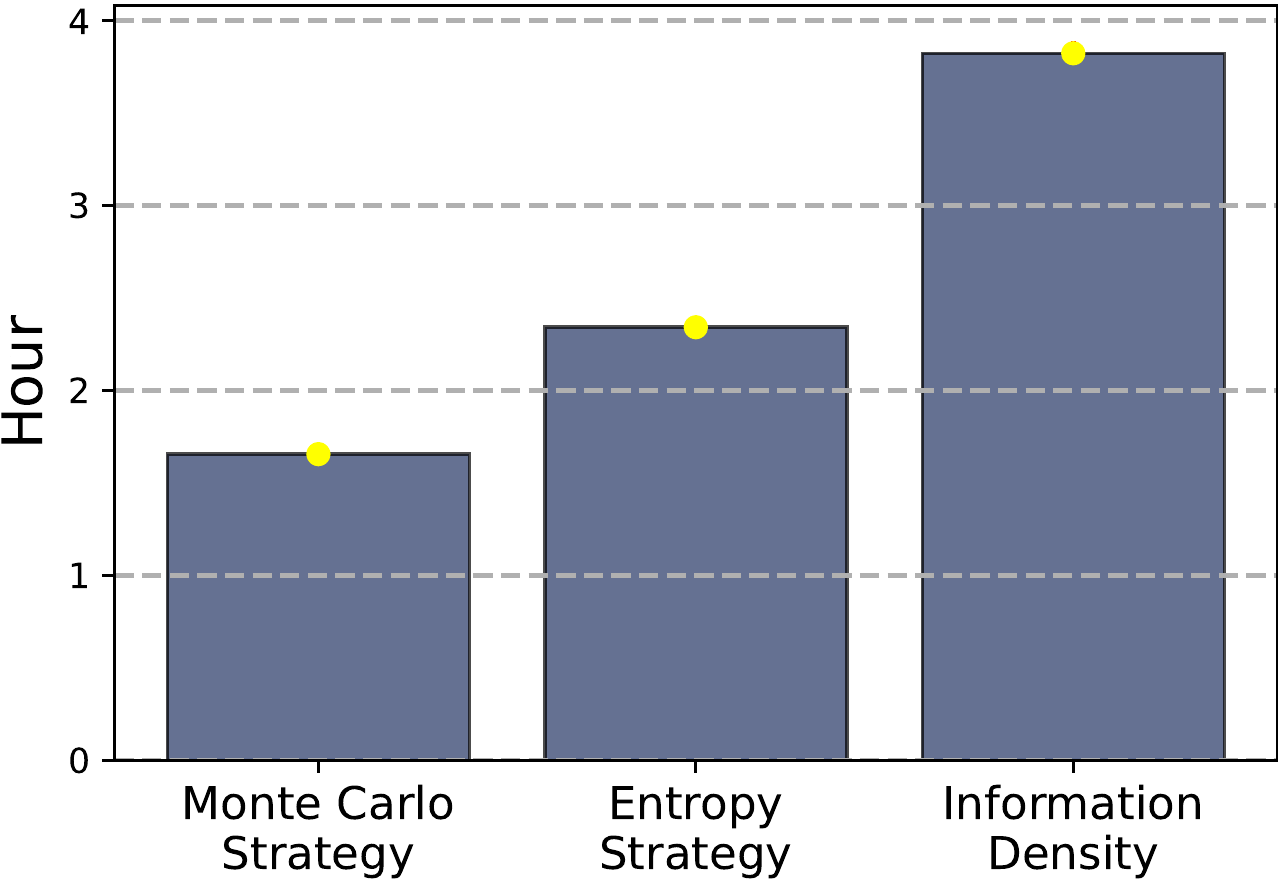}
(b)
\end{minipage}\hfill
\caption{Active learning algorithms are compared in terms of performance on the left and computational costs on the right.}
\label{fig:AL_perf}
\end{figure}

\textbf{Performance Comparison}: Three algorithms are proposed and modified from Algorithm \ref{algo:AL} as follows. 1) Takes a Monte Carlo strategy $\phi^{U}$ in line 5 and there is no need for training in line 3, 2) takes an entropy strategy $\phi^{Ent}$ in line 5 and two layer fully-connected neural networks in line 3 for training, and 3) takes an information density strategy $\phi^{ID}$ in line 5 and two layer fully-connected neural networks in line 3 for training. In all three algorithms, take constants $B = 2^{5}$ and $K = 2^{8}$. To compare performance, run $20$ separate episodes to get enough samples for performance comparisons.

Scenarios consist of both EV and PV integration so we consider the maximum of combined EV and PV to compare performance of algorithms, i.e., the largest sum of the total number of EVs and the total capacity (generation in kW) of PVs. The mean and standard deviation of 20 episodes and the maximum possible hosting capacity in $\mathcal{X}_{1}$ are shown in Fig. \ref{fig:AL_perf} (a) for three algorithms. Observe that information density strategy $\phi^{ID}$ generally outperforms the other two strategies.

\begin{figure*}[hbt!]
\begin{minipage}{.32\linewidth}
\centering
\includegraphics[width=1\linewidth]{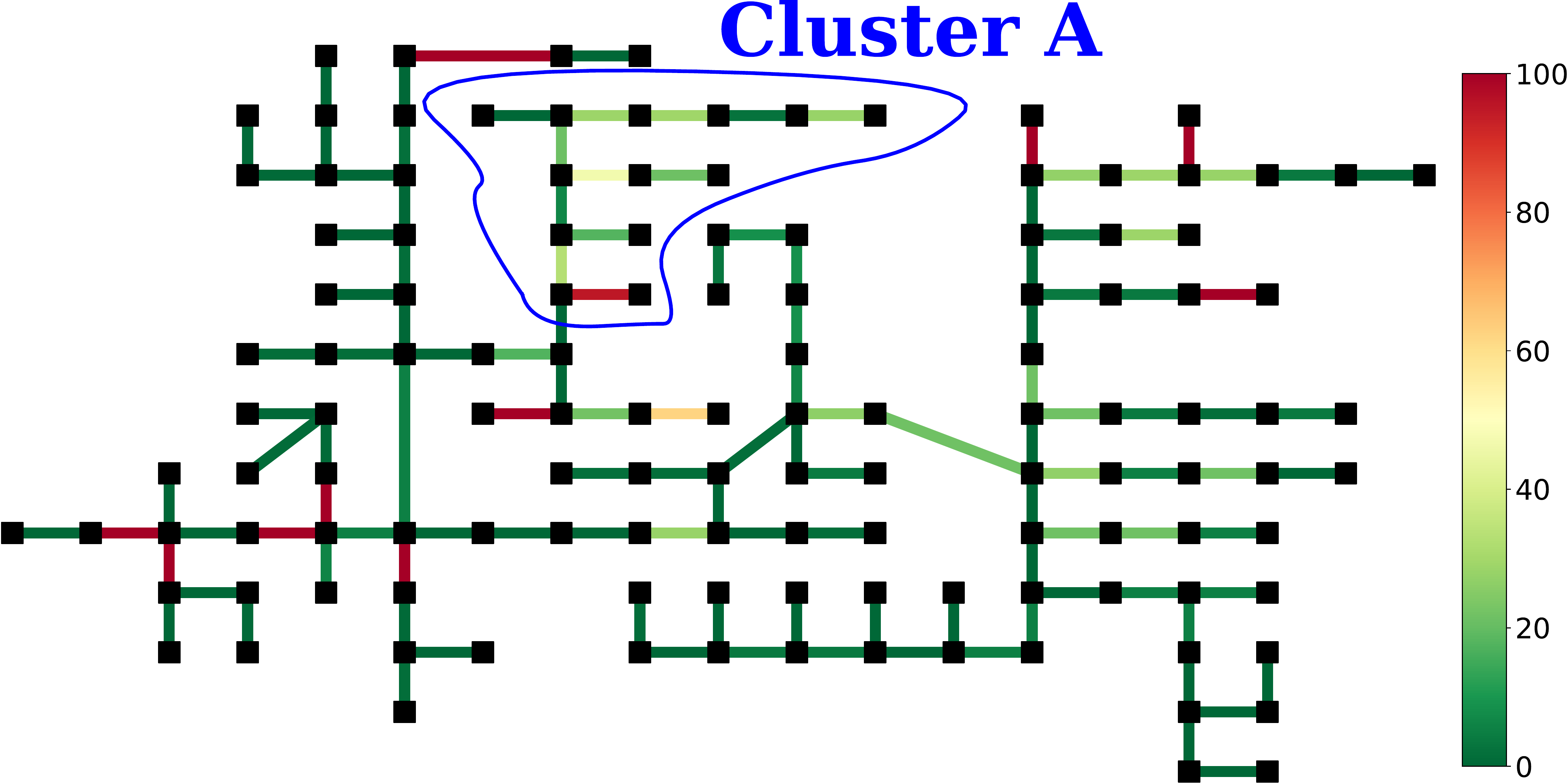}
(a)
\end{minipage}\hfill
\begin{minipage}{.32\linewidth}
\centering
\includegraphics[width=1\linewidth]{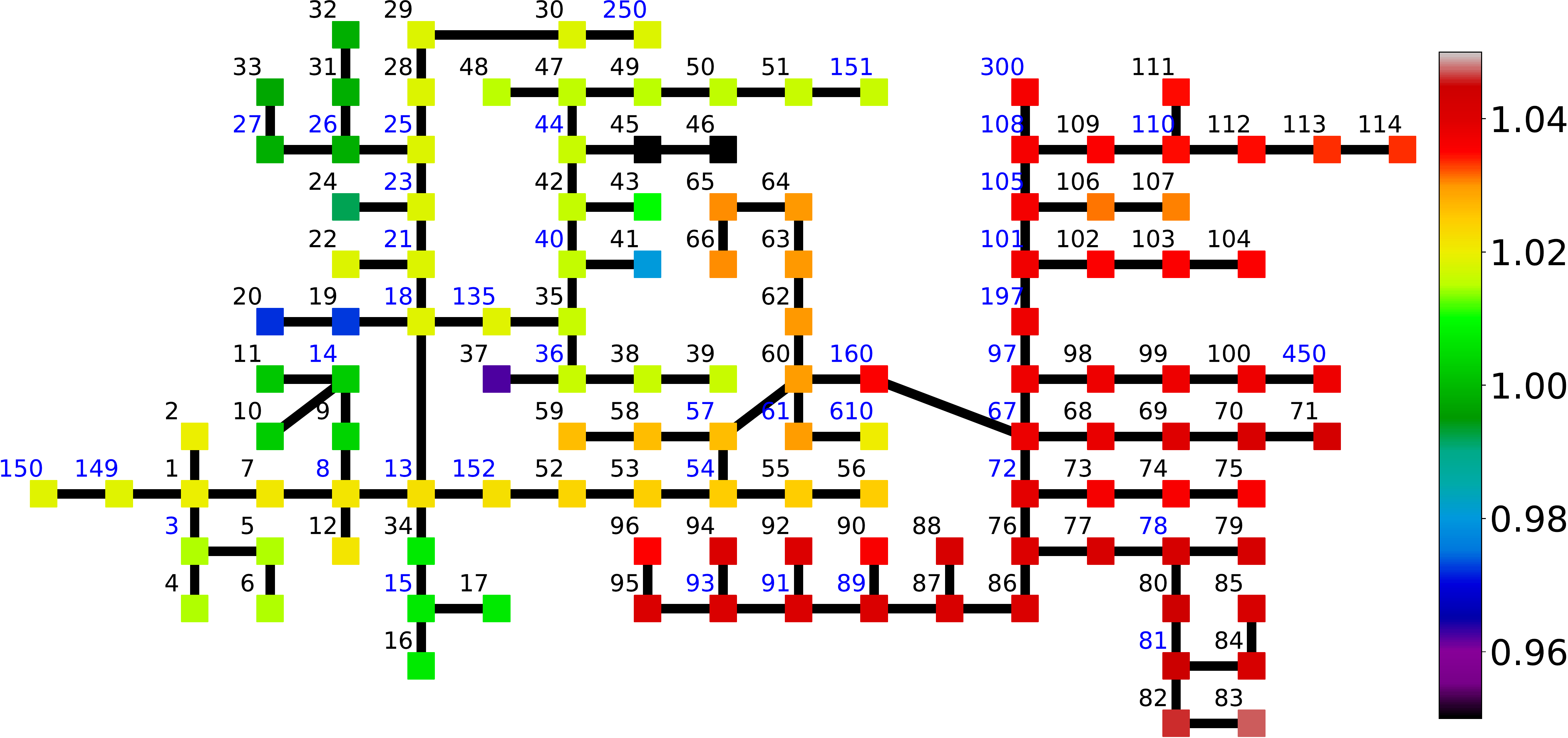}
(b)
\end{minipage}\hfill
\begin{minipage}{.32\linewidth}
\centering
\includegraphics[width=1\linewidth]{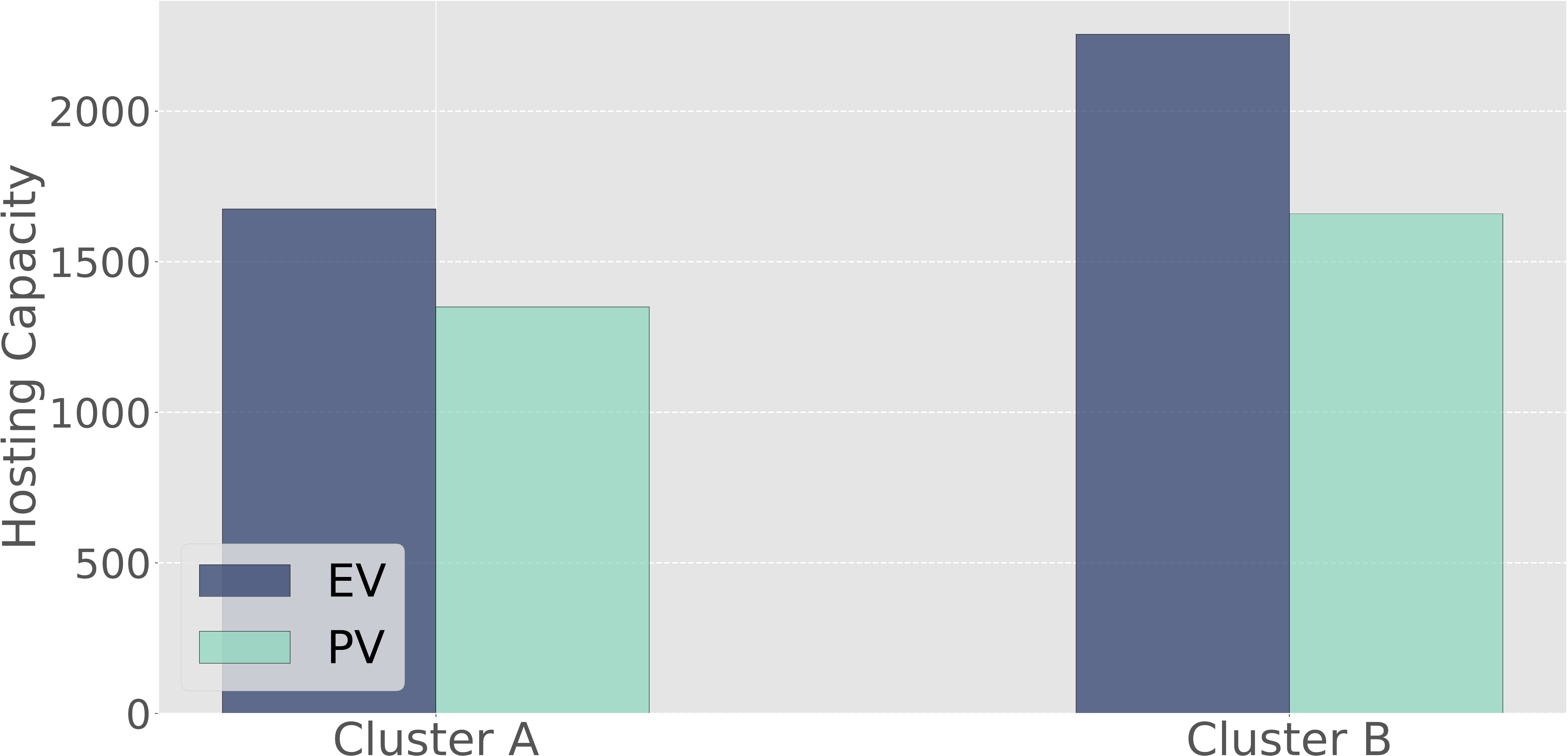}
(c)
\end{minipage}\hfill
\begin{minipage}{.32\linewidth}
\centering
\includegraphics[width=1\linewidth]{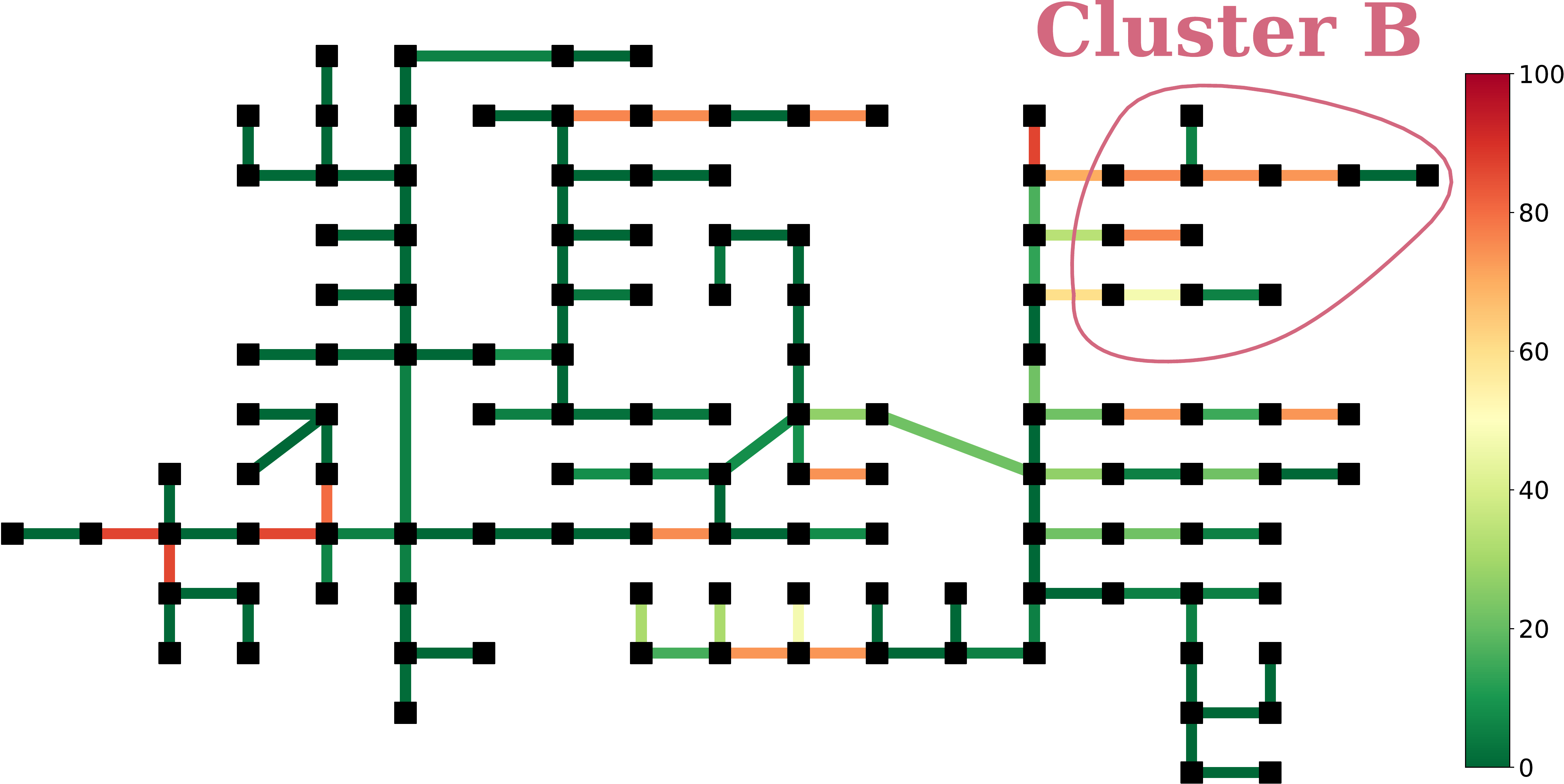}
(d)
\end{minipage}\hfill
\begin{minipage}{.32\linewidth}
\centering
\includegraphics[width=1\linewidth]{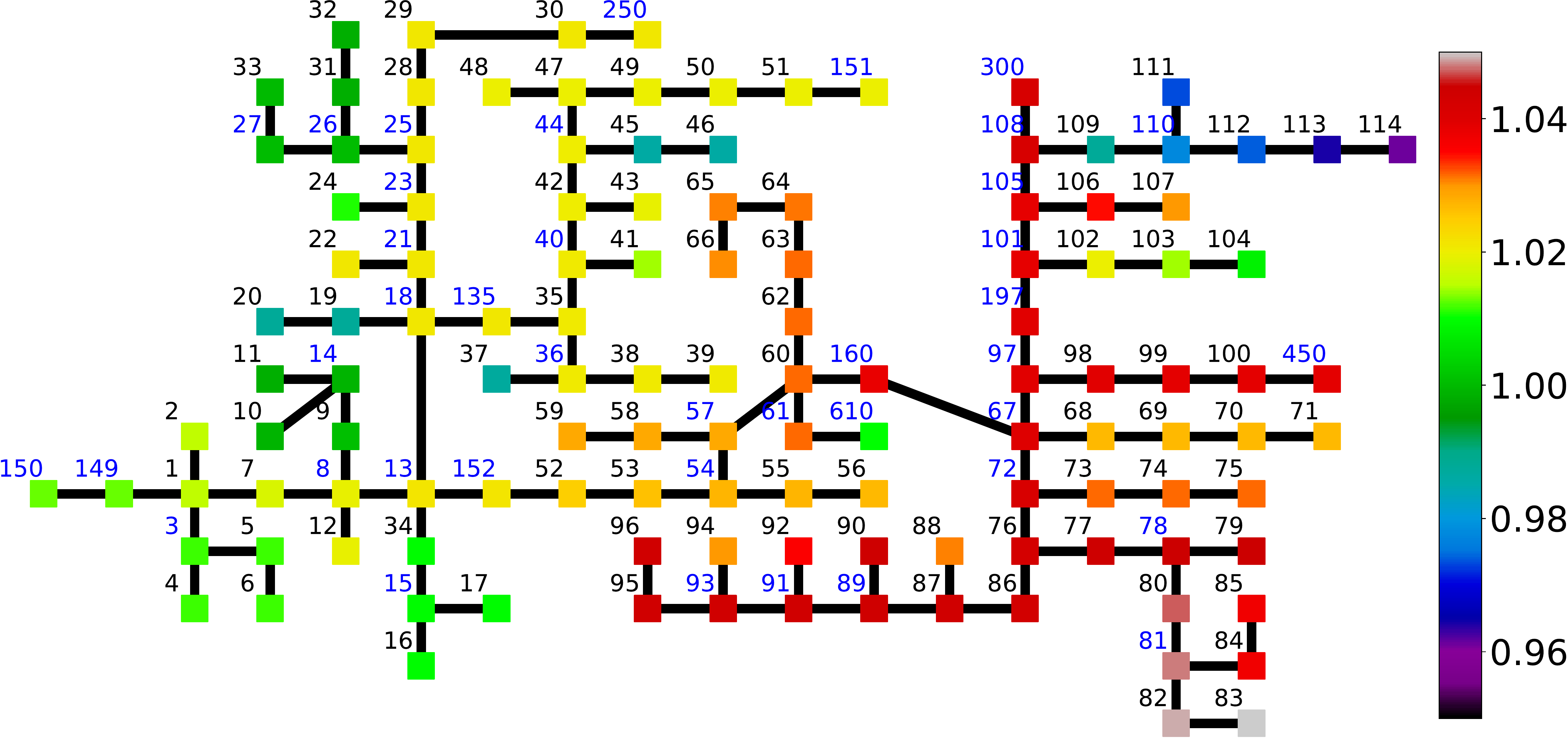}
(e)
\end{minipage}\hfill
\begin{minipage}{.32\linewidth}
\centering
\includegraphics[width=1\linewidth]{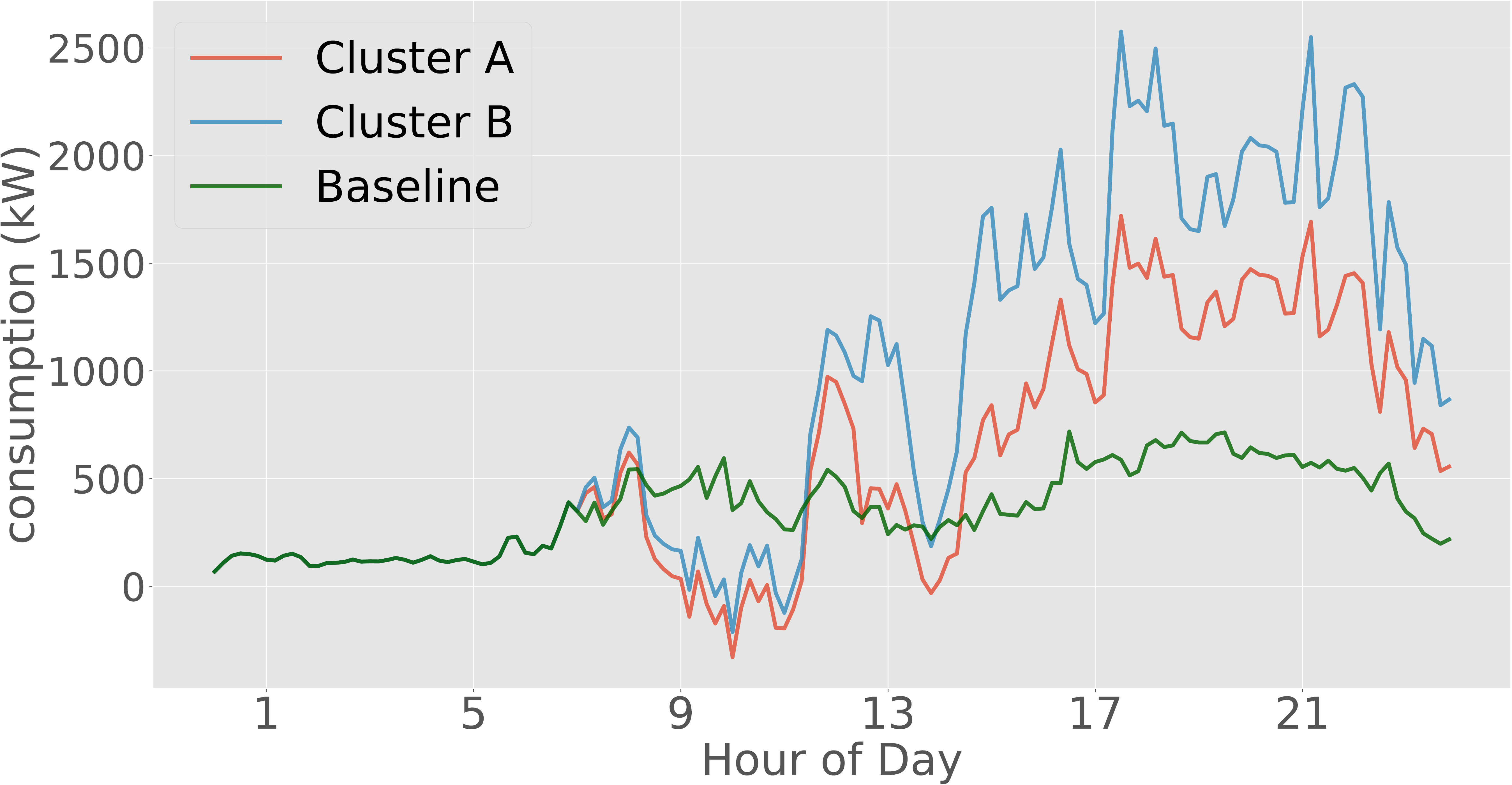}
(f)
\end{minipage}\hfill
\caption{Two regions of Clusters A and B are shown in (a) and (d) respectively and clustering effects of EVs are illustrated. (a) and (b) show line Loading and nodal voltages respectively with clustering effects in Cluster A. (d) and (e) show line Loading and nodal voltages respectively with clustering effects in Cluster B. Corresponding HCs and aggregate system loads are shown in (c) and (f) respectively.}
\label{fig:CE}
\end{figure*}

Total computational costs (in time) are shown in Fig. \ref{fig:AL_perf} (b). Monte Carlo, entropy, information density strategies take 1.7, 2.3, and 3.8 hours respectively. Monte Carlo strategy has no computational overhead due to its simplicity while entropy strategy and information density strategy have computational overhead due to neural networks training and corresponding entropy and information density computation. All algorithms have negligible standard deviations in computational time. Due to its computational simplicity, we take Monte Carlo strategy in the rest of the simulation.

Lastly, we comment about the comparison with centralized optimization-based HCA frameworks. Recall that the objective of the proposed HCA is to include the two factors (a) and (b) whereas optimization-based HCA frameworks find a global optimal solution. Also, recall that adoption patterns of DERs could be far from the global optimal solution when DERs are non-centrally integrated. Thus, it is unclear to make a fair comparison for the two approaches.

\subsection{Clustering Effects of EVs}

In principle, EVs can exist at any location in electric distribution grids after installing a charging station. However, today EVs are quickly adopted in some towns and are not adopted in other towns, i.e., adoption of alternative vehicles such as EVs may be characterized by significant clustering effects correlated with household locations driven by socio-economic and behavioral decisions \cite{muratori2018impact}. While accurate representation for a region of a cluster requires additional domain knowledge (such as transportation and demographics information), we pick two representative clusters to understand how clustering effects of EVs change HC.

\textbf{Design of Scenarios}: Two clusters, Clusters A and B, are arbitrarily picked in IEEE 123-node test feeder network and shown in Fig. \ref{fig:CE} (a) and (d), and two sets of scenarios $\mathcal{X}_{A}$ and $\mathcal{X}_{B}$ are proposed where EV adoption is restricted to the cluster, i.e., EVs exist only in the cluster. In both $\mathcal{X}_{A}$ and $\mathcal{X}_{B}$, assume that there are the same EV charging (consumption) behavior $N_{\text{EV\_TYPE}} = 20$ and the number of EVs is drawn according to the Poisson distribution with the same parameter $\lambda$ uniformly in the cluster. That is, the only difference in $\mathcal{X}_{A}$ and $\mathcal{X}_{B}$ is that EVs are restricted to exist in different regions of clusters.

\textbf{Characteristics of HC}: The results indicate that Clusters A and B induce different operational violations and that Clusters A and B have largely different HC values.

With the clustering effects of EVs in Cluster A, line loading and nodal voltages for a representative scenario in $\mathcal{X}_{A}$ are shown in Fig. \ref{fig:CE} (a) and (b) respectively. Cluster A causes an undervoltage problem (voltage levels below 0.95 p.u.) in buses 45 and 46. Also, there are line loading violations (above 100\% of normal rating) in several lines for peak loads. With the clustering effects of EVs in Cluster B, line loading and nodal voltages for a representative scenario in $\mathcal{X}_{B}$ are shown in Fig. \ref{fig:CE} (d) and (e) respectively. Observe that Cluster B causes an overvoltage problem (voltage levels above 1.05 p.u.) in buses 82 and 83 for peak loads and undervoltage close to violation (voltage level around 0.96 p.u.) in bus 114. Also, line loading is close to violation (i.e., above 90\% of normal rating), and line loading pattern in (d) is very different from (a) except for lines close to the substation.

Moreover, the case study highlights that the aggregate system load (summed over all buses for each time $t \in [T]$) alone is not sufficient to understand the HC and that locations of net injection from DERs are crucial as illustrated in this case study. The same grid can host approximately 2,300 EVs with clustering effects in Cluster B and can host approximately 1,600 EVs with clustering effects in Cluster B as shown in Fig. \ref{fig:CE} (c). Similarly, Fig. \ref{fig:CE} (f) shows Clusters A and B have noticeably different aggregate system loads that the same grid can support. Therefore, we conclude that adoption patterns of DERs that depend on socio-economic behavior play a significant role in the HC limits. 

\begin{figure*}[ht]
\centering
\begin{minipage}[t]{.32\linewidth}
\centering
\includegraphics[width=1\linewidth]{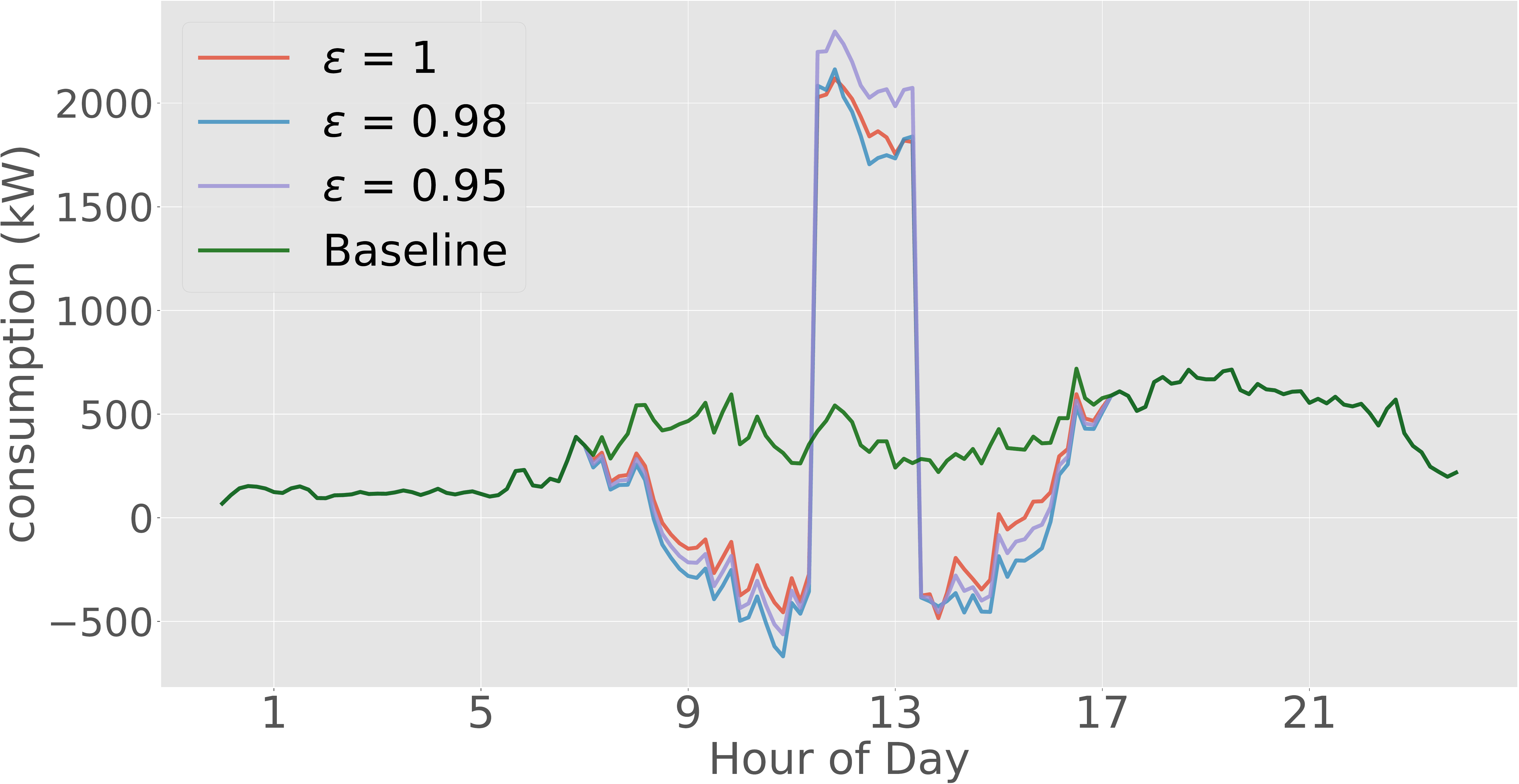}
(a) Uncoordinated
\end{minipage}\hfill
\begin{minipage}[t]{.32\linewidth}
\centering
\includegraphics[width=1\linewidth]{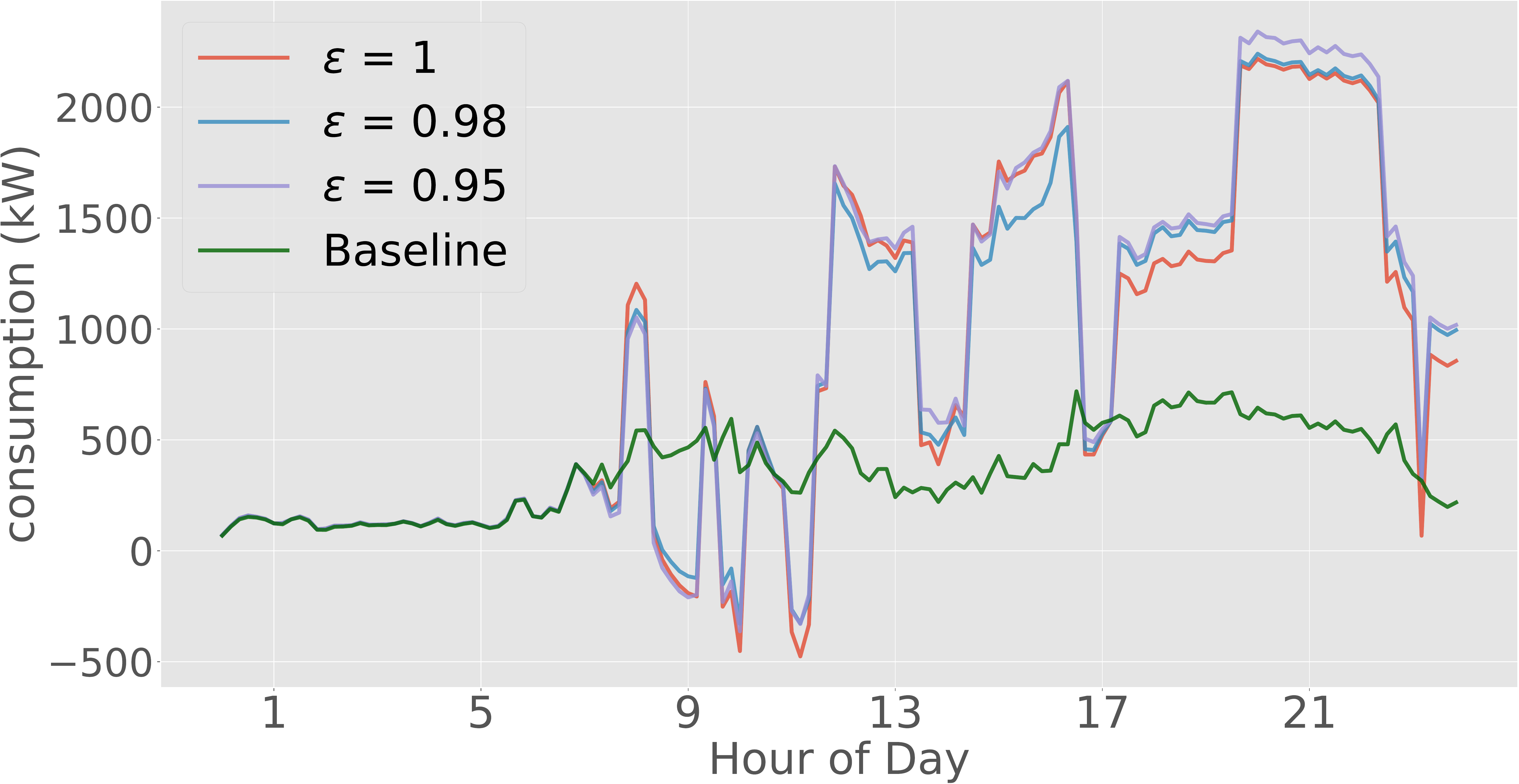}
(b) Mildly coordinated
\end{minipage}\hfill
\begin{minipage}[t]{.32\linewidth}
\centering
\includegraphics[width=1\linewidth]{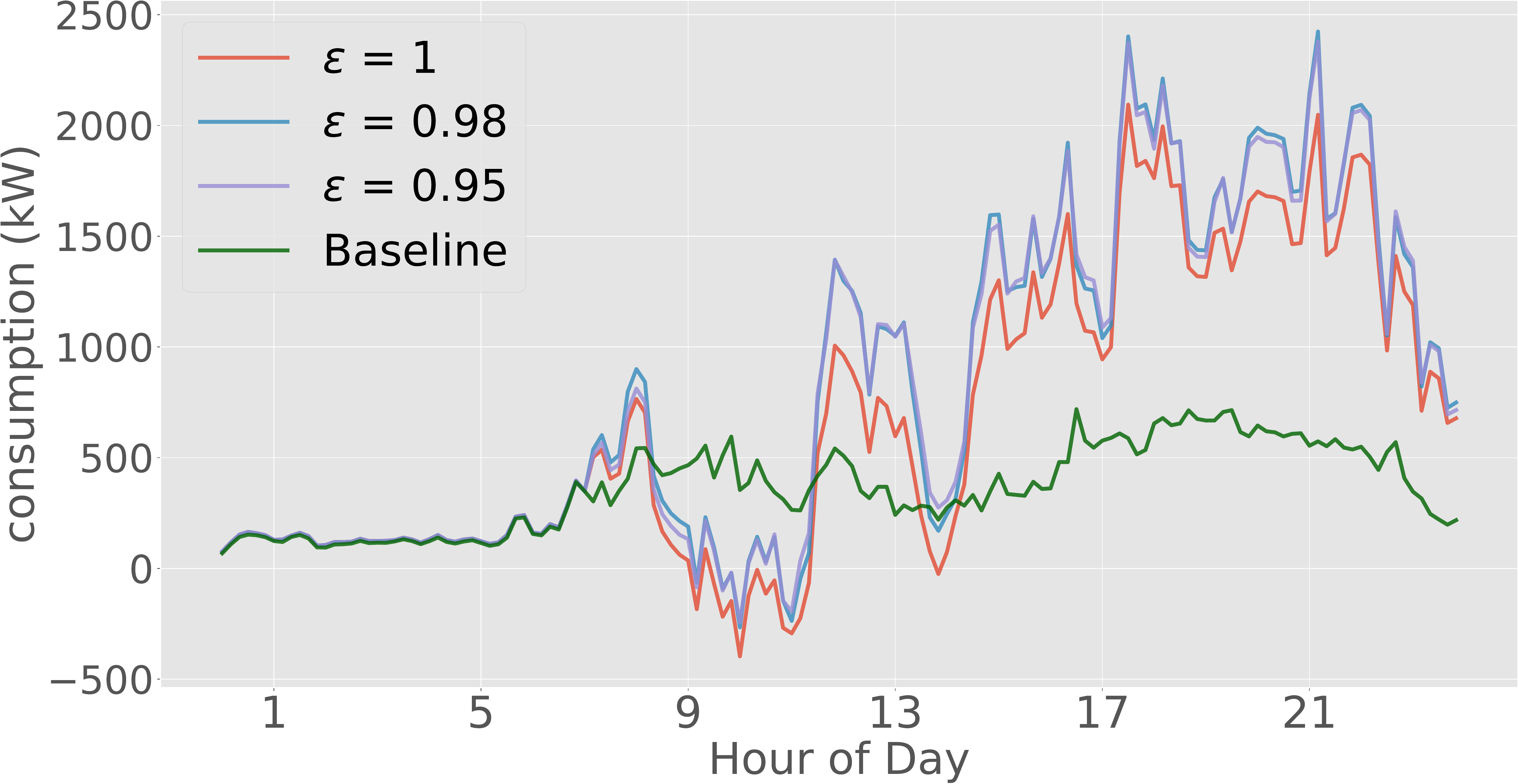}
(c) Moderately coordinated
\end{minipage}\hfill
\caption{Aggregate system consumption (summed over all buses for each time step) is shown for a day for Uncoordinated, Mildly Coordinated, Moderately Coordinated scenarios respectively. Green line shows a baseline consumption in kW without DERs. Observe that additional loads are dispersed as EVs coordinate more. The role of $\bar{\epsilon}$ is illustrated for three values $\bar{\epsilon} = 1, 0.98, 0.95$.}
\label{fig:coordination}
\end{figure*}

\subsection{Degrees of Coordination of EVs}
Different degrees of coordination of EVs are studied for HCA ranging from uncoordinated to moderately coordinated EVs in terms of charging (consumption) behaviors. All EV profiles are data-driven and coordination of EVs is considered.

\begin{figure}[hbt!]
\begin{minipage}[t]{.49\linewidth}
\centering
\includegraphics[width=1\linewidth]{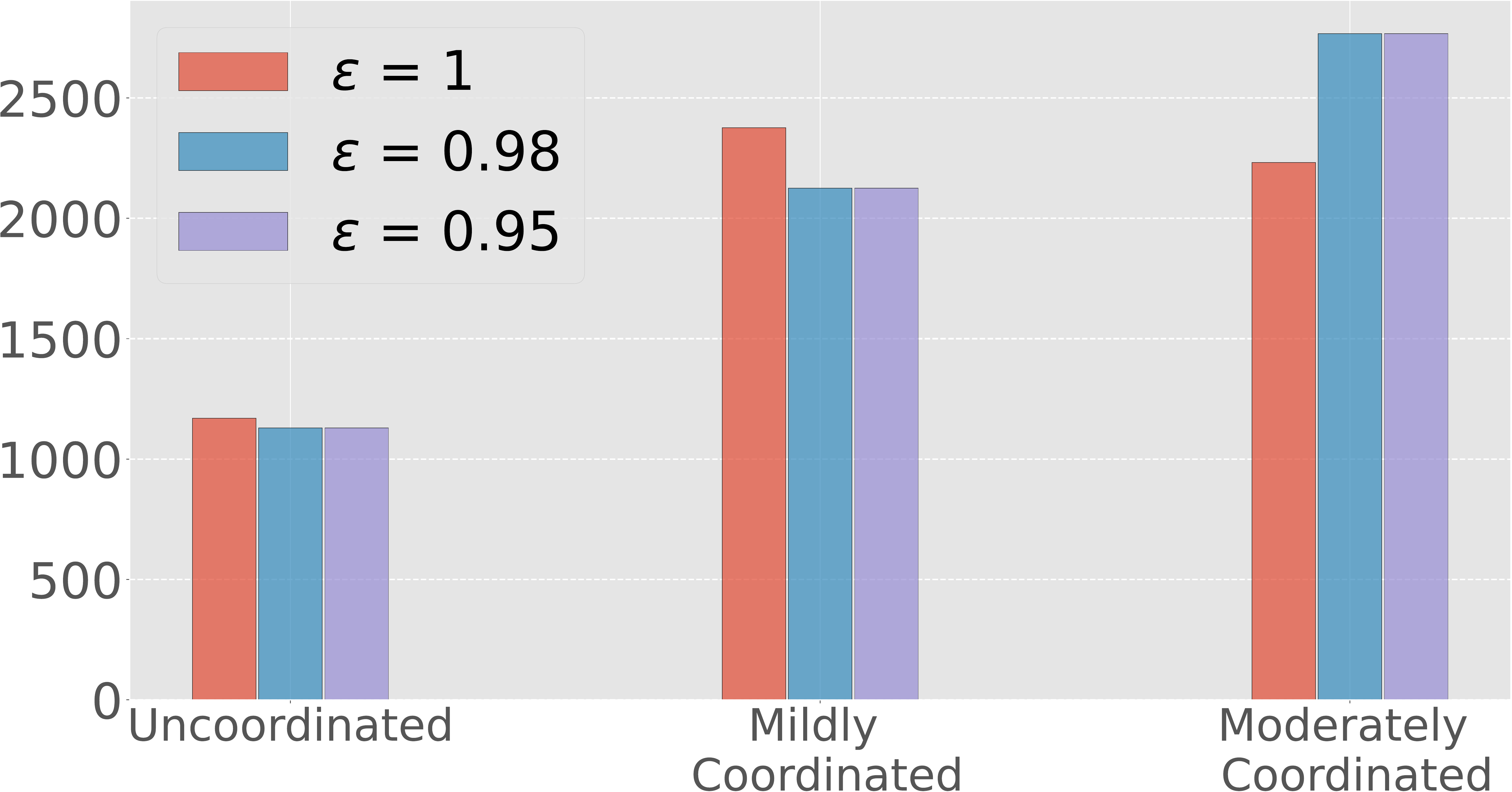}
(a) Hosting Capacity for EV
\end{minipage}\hfill
\begin{minipage}[t]{.49\linewidth}
\centering
\includegraphics[width=1\linewidth]{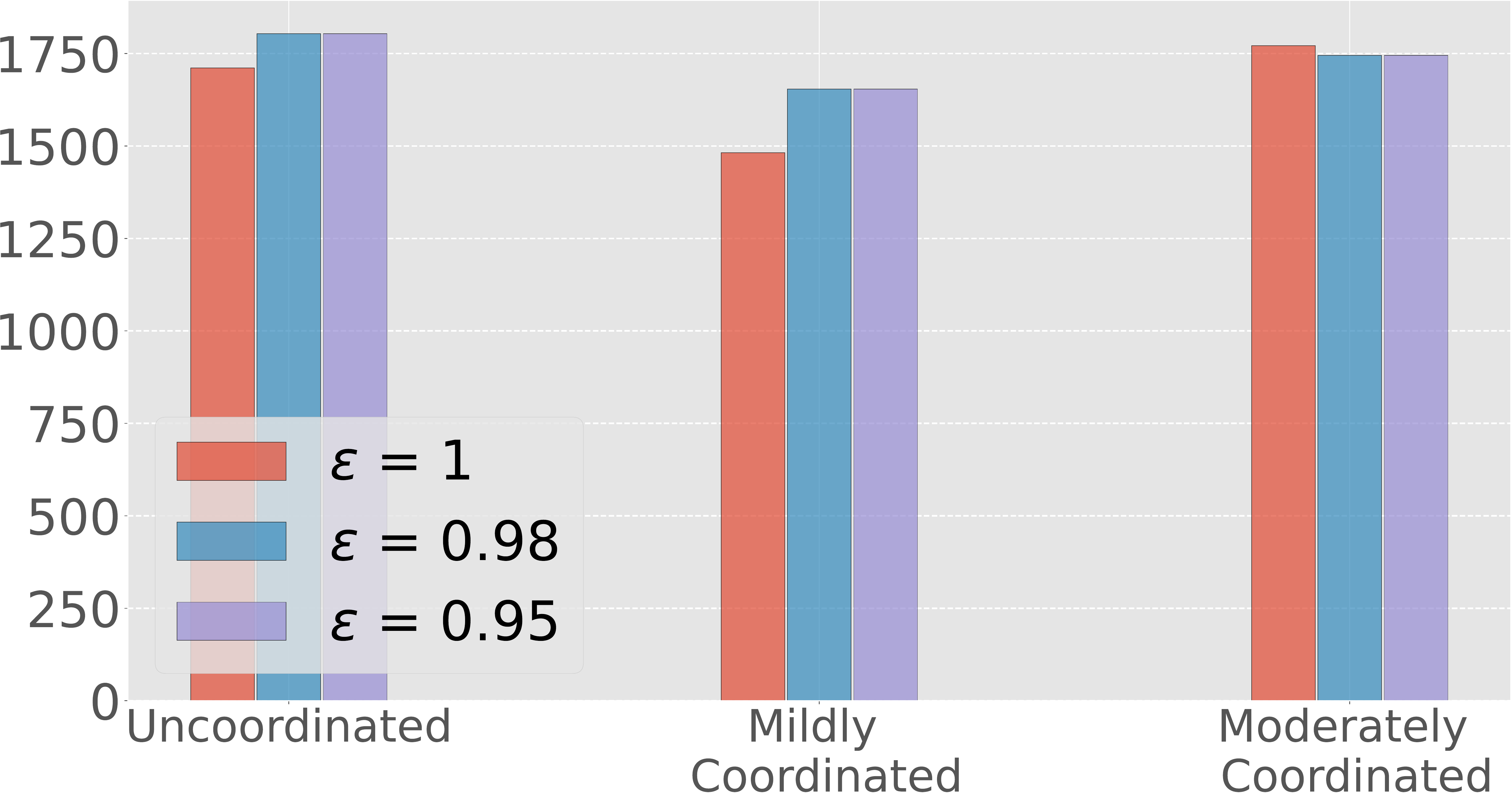}
(b) Hosting Capacity for PV
\end{minipage}\hfill
\caption{Hosting Capacity for EV and PV in uncoordinated, mildly and moderately coordinated scenarios. The role of $\bar{\epsilon} = 1, 0.98, 0.95$ is illustrated.}
\label{fig:cor_hc}
\end{figure}

\textbf{Design of Scenarios}: Three sets of scenarios $\mathcal{X}_{2}, \mathcal{X}_{3}, \mathcal{X}_{4}$ are proposed and called uncoordinated, mildly coordinated and moderately coordinated respectively. The three sets are distinguished in the sense that EVs have different charging (consumption) behaviors as follows. Scenarios $\mathcal{X}_{2}$ are called uncoordinated in a sense that all EVs have identical charging behavior $N_{\text{EV\_TYPE}} = 1$, i.e., all EVs start and stop charging at the same time. Scenarios $\mathcal{X}_{3}$ and $\mathcal{X}_{4}$ are called mildly coordinated and moderately coordinated in a sense that EVs have different charging behavior $N_{\text{EV\_TYPE}} = 5$ and $20$ respectively, i.e., EVs do not necessarily start and stop charging at the same time. Observe that the proposed data-driven EV scenarios have increasing degrees of coordination with $N_{\text{EV\_TYPE}}$. In the three sets, assume the number of EVs (positive integer) is drawn according to the Poisson distribution with a uniform parameter $\lambda$ in all buses. In addition, we examine three probabilistic lower bounds $\bar{\epsilon} = 1, 0.98, 0.95$.

\textbf{Characteristics of HC}: The results illustrate that the same grid can host significantly more EVs when they are coordinated compared to when uncoordinated. Hosting capacity for EVs is shown in Fig. \ref{fig:cor_hc} (a) and it illustrates that the same grid can host approximately 1,000, 2,200, 2,700 EVs when they are uncoordinated, mildly coordinated, moderately coordinated respectively. Hosting capacity for PV is shown in Fig. \ref{fig:cor_hc} (b) and it shows HC mostly remains the same for all three scenarios. Thus, we conclude that HC is noticeably large when EVs are coordinated. Although the results are naturally predicted, the results are worthwhile because it does not explicitly include direct interventions (control) of DERs. In other words, the results are naturally observed from the data and it does not explicitly include system and market interventions. For example, when all EVs are scheduled to follow peak shaving and valley filling programs \cite{ma2011decentralized} that they typically charge EVs during 12am - 6am, i.e., when EVs are disrupted by the intervention, it is also naturally predicted that the same grid can host more EVs. However, it is unclear today whether these grid supporting services and programs are profitable and how many EVs are attracted and will remain attracted to such programs in electric distribution grids. Observe that this is another open and decisive parameter for HC that will definitely vary in towns and cities.

Aggregate system loads (summed over all buses for each time $t \in [T]$) are illustrated in Fig. \ref{fig:coordination} for representative scenarios in $\mathcal{X}_{2}, \mathcal{X}_{3}, \mathcal{X}_{4}$, respectively. Baseline (green line) represents default daily aggregate system loads (consumption in kW) without DERs. There is a big irregular jump because EVs are uncoordinated (start and stop charging at the same time), and additional loads from EVs are dispersed and there is more consumption during peak hours. Also, the aggregate system loads are occasionally below 0 during the daytime when the total renewable generation exceeds the aggregate system loads.

Lastly, the role of probabilistic lower bound $\bar{\epsilon}$ in Eq. \eqref{eq:HC_SA} is investigated that three values of $\bar{\epsilon} = 1, 0.98, 0.95$ are considered for $\mathcal{X}_{2}, \mathcal{X}_{3}, \mathcal{X}_{4}$. Observe in Fig. \ref{fig:coordination} and \ref{fig:cor_hc} that $\bar{\epsilon}$ slightly affects the system performance for HCA only marginal changes are observed for different values of $\bar{\epsilon}$ that the aggregate system consumption and HC limits are marginally changed for different values of $\bar{\epsilon}$.

\subsection{A Combined Integration of PVs and EVs}

A combined integration of PVs and EVs is investigated to understand synergies between PVs and EVs. Our creative imaginations permit that there can be a few directions for a combination of PVs and EVs in electric distribution grids, however, we consider a simple combination for the scope of this work.

\textbf{Design of Scenarios}: Two sets of scenarios $\mathcal{X}_{5}$ and $\mathcal{X}_{6}$ are proposed and designed as follows. Assume that the number of EVs (positive integer) is drawn according to the Poisson distribution with a uniform parameter $\lambda$ in all buses and that charging behaviors $N_{\text{EV\_TYPE}} = 20$ in $\mathcal{X}_{5}$ and $\mathcal{X}_{6}$. On the other hand, for all scenarios in $\mathcal{X}_{5}$, consider that each EV is co-located with a PV in the same bus whereas there is no PV for all scenarios in $\mathcal{X}_{6}$. With this simple setup, we consider whether the same grid can host more EVs or not.

\textbf{Characteristics of HC}: The results indicate that a combined integration of EVs and PVs does not have significant adverse or synergy. To be specific, the same grid can host approximately 2,800 and 2,900 EVs for scenarios in $\mathcal{X}_{5}$ and $\mathcal{X}_{6}$ respectively. Also, hosting capacity for PV is approximately 1,800 for scenarios in $\mathcal{X}_{5}$. The results can be explained from Fig. \ref{fig:coordination}. In Fig. \ref{fig:coordination} (a), notice that PVs have a significant injection that the aggregate system consumption is below zero during the peak hours for PVs (between 10AM and 3PM) whereas the injection is almost negligible after 4pm. On the other hand, observe in Fig. \ref{fig:coordination} (c) that the peak hours for EVs are after 5PM. Thus, we conclude that interactions of two technologies (PVs and EVs) are not significant. We remark that the proposed method is data-driven so it is possible that the results may be slightly (but not significantly) different when data sources for PVs and EVs are different. For example, DER profiles could change for summer and winter periods. Thus, it is expected that HC values do slightly change but may not significantly change.

%% file: 9-Conclusion.tex
\section{Concluding Remarks}\label{sec:conc}

This work presents a HCA framework with two factors in mind: (a) adoption patterns of DERs driven by socio-economic behaviors and (b) how DERs are controlled and managed. To address the challenge, a data-driven HCA framework and active learning are proposed to effectively explore scenarios. We illustrate the properties of active learning and characteristics of HC in a 3-bus network example, which illustrates that there are numerous scenarios between feasible and infeasible sets. This poses challenges from a system operator's perspective because it is unclear how to interpret numerous feasible scenarios. On the other hand, when additional domain knowledge about (a) and (b) is available, we show that it is possible to consider a few most insightful scenarios and not worry about all others. Next, on a larger network, we present the performance of active learning algorithms and consider detailed case studies about (a) and (b). Simulation results highlight that the HC and its interpretation significantly change subject to (a) and (b). Future work will investigate further improvement on the region estimate of the HC, as well as the public policy implication of the HCA for DER planning purposes.

%% file: main.bbl
\begin{thebibliography}{10}
\providecommand{\url}[1]{#1}
\csname url@samestyle\endcsname
\providecommand{\newblock}{\relax}
\providecommand{\bibinfo}[2]{#2}
\providecommand{\BIBentrySTDinterwordspacing}{\spaceskip=0pt\relax}
\providecommand{\BIBentryALTinterwordstretchfactor}{4}
\providecommand{\BIBentryALTinterwordspacing}{\spaceskip=\fontdimen2\font plus
\BIBentryALTinterwordstretchfactor\fontdimen3\font minus
  \fontdimen4\font\relax}
\providecommand{\BIBforeignlanguage}[2]{{%
\expandafter\ifx\csname l@#1\endcsname\relax
\typeout{** WARNING: IEEEtran.bst: No hyphenation pattern has been}%
\typeout{** loaded for the language `#1'. Using the pattern for}%
\typeout{** the default language instead.}%
\else
\language=\csname l@#1\endcsname
\fi
#2}}
\providecommand{\BIBdecl}{\relax}
\BIBdecl

\bibitem{joskow2020transmission}
P.~L. Joskow, ``Transmission capacity expansion is needed to decarbonize the
  electricity sector efficiently,'' \emph{Joule}, vol.~4, no.~1, pp. 1--3,
  2020.

\bibitem{xie2021toward}
L.~Xie, C.~Singh, S.~K. Mitter, M.~A. Dahleh, and S.~S. Oren, ``Toward
  carbon-neutral electricity and mobility: Is the grid infrastructure ready?''
  \emph{Joule}, vol.~5, no.~8, pp. 1908--1913, 2021.

\bibitem{ebad2016approach}
M.~Ebad and W.~M. Grady, ``An approach for assessing high-penetration pv impact
  on distribution feeders,'' \emph{Electric Power Systems Research}, vol. 133,
  pp. 347--354, 2016.

\bibitem{mohammadi2016challenges}
P.~Mohammadi and S.~Mehraeen, ``Challenges of pv integration in low-voltage
  secondary networks,'' \emph{IEEE Transactions on Power Delivery}, vol.~32,
  no.~1, pp. 525--535, 2016.

\bibitem{gaunt2017voltage}
C.~Gaunt, E.~Namanya, and R.~Herman, ``Voltage modelling of lv feeders with
  dispersed generation: Limits of penetration of randomly connected
  photovoltaic generation,'' \emph{Electric Power Systems Research}, vol. 143,
  pp. 1--6, 2017.

\bibitem{leemput2014impact}
N.~Leemput, F.~Geth, J.~Van~Roy, A.~Delnooz, J.~B{\"u}scher, and J.~Driesen,
  ``Impact of electric vehicle on-board single-phase charging strategies on a
  flemish residential grid,'' \emph{IEEE Transactions on Smart Grid}, vol.~5,
  no.~4, pp. 1815--1822, 2014.

\bibitem{linna2017congestion}
N.~Linna, W.~Fushuan, L.~Weijia, M.~Jinling, L.~Guoying, and D.~Sanlei,
  ``Congestion management with demand response considering uncertainties of
  distributed generation outputs and market prices,'' \emph{Journal of Modern
  Power Systems and Clean Energy}, vol.~5, no.~1, pp. 66--78, 2017.

\bibitem{johansson2019investigation}
S.~Johansson, J.~Persson, S.~Lazarou, and A.~Theocharis, ``Investigation of the
  impact of large-scale integration of electric vehicles for a swedish
  distribution network,'' \emph{Energies}, vol.~12, no.~24, p. 4717, 2019.

\bibitem{baldenko2016determination}
N.~Baldenko and S.~Behzadirafi, ``Determination of photovoltaic hosting
  capacity on radial electric distribution feeders,'' in \emph{2016 IEEE
  International Conference on Power System Technology (POWERCON)}.\hskip 1em
  plus 0.5em minus 0.4em\relax IEEE, 2016, pp. 1--4.

\bibitem{navarro2015probabilistic}
A.~Navarro-Espinosa and L.~F. Ochoa, ``Probabilistic impact assessment of low
  carbon technologies in lv distribution systems,'' \emph{IEEE Transactions on
  Power Systems}, vol.~31, no.~3, pp. 2192--2203, 2015.

\bibitem{abad2018probabilistic}
M.~S.~S. Abad, J.~Ma, D.~Zhang, A.~S. Ahmadyar, and H.~Marzooghi,
  ``Probabilistic assessment of hosting capacity in radial distribution
  systems,'' \emph{IEEE Transactions on Sustainable Energy}, vol.~9, no.~4, pp.
  1935--1947, 2018.

\bibitem{fatima2020review}
S.~Fatima, V.~P{\"u}vi, and M.~Lehtonen, ``Review on the pv hosting capacity in
  distribution networks,'' \emph{Energies}, vol.~13, no.~18, p. 4756, 2020.

\bibitem{bollen2017hosting}
M.~H. Bollen and S.~K. R{\"o}nnberg, ``Hosting capacity of the power grid for
  renewable electricity production and new large consumption equipment,''
  \emph{Energies}, vol.~10, no.~9, p. 1325, 2017.

\bibitem{paudyal2021ev}
P.~Paudyal, S.~Ghosh, S.~Veda, D.~Tiwari, and J.~Desai, ``Ev hosting capacity
  analysis on distribution grids,'' in \emph{2021 IEEE Power \& Energy Society
  General Meeting (PESGM)}.\hskip 1em plus 0.5em minus 0.4em\relax IEEE, 2021,
  pp. 1--5.

\bibitem{rout2020hosting}
S.~Rout and G.~Biswal, ``Hosting capacity assessment of electric vehicles
  integration in active distribution system,'' in \emph{Journal of Physics:
  Conference Series}, vol. 1478, no.~1.\hskip 1em plus 0.5em minus 0.4em\relax
  IOP Publishing, 2020, p. 012006.

\bibitem{liu2020probabilistic}
D.~Liu, C.~Wang, F.~Tang, and Y.~Zhou, ``Probabilistic assessment of hybrid
  wind-pv hosting capacity in distribution systems,'' \emph{Sustainability},
  vol.~12, no.~6, p. 2183, 2020.

\bibitem{fachrizal2021combined}
R.~Fachrizal, U.~H. Ramadhani, J.~Munkhammar, and J.~Wid{\'e}n, ``Combined
  pv--ev hosting capacity assessment for a residential lv distribution grid
  with smart ev charging and pv curtailment,'' \emph{Sustainable Energy, Grids
  and Networks}, vol.~26, p. 100445, 2021.

\bibitem{e2022combined}
L.~E.~S. e~Silva and J.~P.~A. Vieira, ``Combined pv-pev hosting capacity
  analysis in low-voltage distribution networks,'' \emph{Electric Power Systems
  Research}, vol. 206, p. 107829, 2022.

\bibitem{madavan2022conditional}
A.~N. Madavan, N.~Dahlin, S.~Bose, and L.~Tong, ``Conditional value at
  risk-sensitive solar hosting capacity analysis in distribution networks,''
  \emph{arXiv preprint arXiv:2204.09096}, 2022.

\bibitem{geng2021probabilistic}
X.~Geng, L.~Tong, A.~Bhattacharya, B.~Mallick, and L.~Xie, ``Probabilistic
  hosting capacity analysis via bayesian optimization,'' in \emph{2021 IEEE
  Power \& Energy Society General Meeting (PESGM)}.\hskip 1em plus 0.5em minus
  0.4em\relax IEEE, 2021, pp. 1--5.

\bibitem{taheri2020fast}
S.~Taheri, M.~Jalali, V.~Kekatos, and L.~Tong, ``Fast probabilistic hosting
  capacity analysis for active distribution systems,'' \emph{IEEE Transactions
  on Smart Grid}, vol.~12, no.~3, pp. 2000--2012, 2020.

\bibitem{munikoti2022novel}
S.~Munikoti, M.~Abujubbeh, K.~Jhala, and B.~Natarajan, ``A novel framework for
  hosting capacity analysis with spatio-temporal probabilistic voltage
  sensitivity analysis,'' \emph{International Journal of Electrical Power \&
  Energy Systems}, vol. 134, p. 107426, 2022.

\bibitem{low2014convex}
S.~H. Low, ``Convex relaxation of optimal power flow—part i: Formulations and
  equivalence,'' \emph{IEEE Transactions on Control of Network Systems},
  vol.~1, no.~1, pp. 15--27, 2014.

\bibitem{baran1989network}
M.~E. Baran and F.~F. Wu, ``Network reconfiguration in distribution systems for
  loss reduction and load balancing,'' \emph{IEEE Power Engineering Review},
  vol.~9, no.~4, pp. 101--102, 1989.

\bibitem{geng2019data}
X.~Geng and L.~Xie, ``Data-driven decision making in power systems with
  probabilistic guarantees: Theory and applications of chance-constrained
  optimization,'' \emph{Annual reviews in control}, vol.~47, pp. 341--363,
  2019.

\bibitem{muratori2018impact}
M.~Muratori, ``Impact of uncoordinated plug-in electric vehicle charging on
  residential power demand,'' \emph{Nature Energy}, vol.~3, no.~3, pp.
  193--201, 2018.

\bibitem{settles2009active}
B.~Settles, ``Active learning literature survey,'' \emph{Computer Sciences
  Technical Report 1648, University of Wisconsin-Madison}, 2009.

\bibitem{karamcheti2021mind}
S.~Karamcheti, R.~Krishna, L.~Fei-Fei, and C.~D. Manning, ``Mind your outliers!
  investigating the negative impact of outliers on active learning for visual
  question answering,'' \emph{arXiv preprint arXiv:2107.02331}, 2021.

\bibitem{OpenDSS}
R.~C. Dugan and D.~Montenegro, ``{The Open Distribution System
  Simulator™(OpenDSS), Reference Guide},'' \emph{Electric Power Research
  Institute (EPRI)}, 2018.

\bibitem{OpenDSSDirect}
D.~Krishnamurthy, ``{OpenDSSDirect},''
  \url{https://dss-extensions.org/OpenDSSDirect.py/index.html}, 2017.

\bibitem{zimmerman2010matpower}
R.~D. Zimmerman, C.~E. Murillo-S{\'a}nchez, and R.~J. Thomas, ``Matpower:
  Steady-state operations, planning, and analysis tools for power systems
  research and education,'' \emph{IEEE Transactions on power systems}, vol.~26,
  no.~1, pp. 12--19, 2010.

\bibitem{shalev2014understanding}
S.~Shalev-Shwartz and S.~Ben-David, \emph{Understanding machine learning: From
  theory to algorithms}.\hskip 1em plus 0.5em minus 0.4em\relax Cambridge
  university press, 2014.

\bibitem{lew2013western}
D.~Lew, G.~Brinkman, E.~Ibanez, A.~Florita, M.~Heaney, B.-M. Hodge, M.~Hummon,
  G.~Stark, J.~King, S.~A. Lefton \emph{et~al.}, ``Western wind and solar
  integration study phase 2,'' National Renewable Energy Lab.(NREL), Golden, CO
  (United States), Tech. Rep., 2013.

\bibitem{ma2011decentralized}
Z.~Ma, D.~S. Callaway, and I.~A. Hiskens, ``Decentralized charging control of
  large populations of plug-in electric vehicles,'' \emph{IEEE Transactions on
  control systems technology}, vol.~21, no.~1, pp. 67--78, 2011.

\end{thebibliography}
